\documentclass[aps,prd,showkeys,preprint,amsmath,amssymb,nofootinbib]{revtex4-1}

\usepackage{comment}
\usepackage{color}
\usepackage{amsmath}
\usepackage{amssymb}
\usepackage{amsthm}
\usepackage{mathrsfs}
\usepackage{graphicx}
\usepackage{marvosym}
\usepackage{amsmath,bm}
\usepackage{array}
\usepackage{latexsym}
\usepackage{enumerate}
\usepackage{slashed}
\usepackage{hyperref}
\usepackage{color}
\usepackage[margin=20pt,small]{caption}
 \usepackage[normalem]{ulem}

\begin{document}
\fontsize{12pt}{12pt}\selectfont
\title{Nonlinear Responses of Chiral Fluids from Kinetic Theory}
\preprint{RIKEN-QHP-260, RIKEN-STAMP-36}

\author{Yoshimasa Hidaka$^{1,2}$, Shi Pu$^{3}$, Di-Lun Yang$^{1}$}
\affiliation{
	$^1$Theoretical Research Division, Nishina Center,
	RIKEN, Wako, Saitama 351-0198, Japan.\\
	$^2$iTHEMS Program, RIKEN, Wako, Saitama 351-0198, Japan.\\
	$^3$	Department of Physics, The University of Tokyo,
	7-3-1 Hongo, Bunkyo-ku, Tokyo 113-0033, Japan.%
	} %
\begin{abstract}
	The second-order nonlinear responses of inviscid chiral fluids near local equilibrium are investigated by applying the chiral kinetic theory (CKT) incorporating side-jump effects.  It is shown that the local equilibrium distribution function can be non-trivially introduced in a co-moving frame with respect to the fluid velocity when the quantum corrections in collisions are involved.
For the study of anomalous transport, 
	contributions from both quantum corrections in anomalous hydrodynamic equations of motion and those from the CKT and Wigner functions are considered under the relaxation-time (RT) approximation, which result in anomalous charge Hall currents propagating along the cross product of the background electric field and the temperature (or chemical-potential) gradient and of the temperature and chemical-potential gradients. On the other hand, the nonlinear quantum correction on the charge density vanishes in the classical RT approximation, which in fact satisfies the matching condition given by the anomalous equation obtained from the CKT.        
\end{abstract}

\maketitle
\section{Introduction}
In recent years, there have been mounting interests in the transport of relativistic Weyl fermions in both nuclear physics and condensed matter systems. One of the renown examples is the study of chiral magnetic effect (CME) associated with the chiral anomaly, from which a vector-charge current is generated along magnetic fields for Weyl fermions,
${\bf J}_V=\sigma_B{\bf B}$, where $\sigma_B$ represents the CME conductivity characterized by chiral imbalance and the coefficient of chiral anomaly \cite{Vilenkin:1980fu}. Such an effect draws much attention for the research in heavy ion collisions, where a system with approximated chiral symmetry could be realized in quark gluon plasmas (QGP). In addition, a strong magnetic field generated by colliding nuclei and the local chiral imbalance stemming from topological excitations makes heavy ion collisions a suitable testing ground for CME \cite{Kharzeev:2007jp,Fukushima:2008xe,Kharzeev:2009pj,Kharzeev:2015znc}. On the other hand, in Weyl semimetals, the quasi-particles at Weyl nodes mimic relativistic Weyl fermions. By pumping a nonzero axial chemical potential through applied electric and magnetic fields parallel to each other, longitudinal negative magneto-resistance associated with CME has been recently observed \cite{Li:2014bha}. Moreover, not only magnetic fields but also vorticity could trigger an anomalous current, known as the chiral vortical effect (CVE) \cite{Vilenkin:1979ui}. As the counter part of vector currents, both magnetic fields and vorticity could also induce axial currents, where the former case is dubbed as chiral separation effect (CSE) \cite{Fukushima:2008xe}. The interplay between CME and CSE could further yield propagating charge-density waves \cite{Kharzeev:2010gd}, which only rely on local fluctuations of vector and axial charges and hence exist even in the absence of net vector/axial chemical potentials. Such charge-density waves known as chiral magnetic waves (CMW) could lead to potentially measurable observables in heavy ion collisions \cite{Burnier:2011bf}.
There are also some studies for axial currents induced by electric fields via interactions \cite{Huang:2013iia,Pu:2014cwa,Pu:2014fva}.     

On the theoretical side, these quantum effects particularly for CME/CVE associated with quantum anomalies have been investigated from various approaches including field theories based on Kubo formula \cite{Fukushima:2008xe,Kharzeev:2009pj,Landsteiner:2011cp}, kinetic theory \cite{Gao:2012ix, Son:2012wh, Stephanov:2012ki, Son:2012zy,Chen:2012ca,Manuel:2013zaa,
Manuel:2014dza,Kharzeev:2016sut, Huang:2017tsq,Gao:2017gfq}, 
relativistic hydrodynamics \cite{Son:2009tf,Neiman:2010zi,Sadofyev:2010pr,Pu:2010as,Kharzeev:2011ds},
lattice simulations \cite{Abramczyk:2009gb,Buividovich:2009wi,Buividovich:2009zzb,Buividovich:2010tn,Yamamoto:2011gk,Mueller:2016ven,Mace:2016shq}, and
gauge/gravity duality \cite{Erdmenger2009,Torabian2009a,Banerjee2011,Landsteiner2011}. Peculiarly, recent progress in chiral kinetic theory (CKT) with the manifestation of Lorenz symmetry and the incorporation of collisions has facilitated our understandings on anomalous transport out of equilibrium \cite{Chen:2014cla,Chen:2015gta,Hidaka:2016yjf}. It is found that the presence of side-jump terms in Wigner functions or equivalently the modified Lorentz transformation of distribution functions pertinent to side-jump phenomena of Weyl fermions is crucial for Lorentz covariance and the contribution to CVE. Regarding the Lorentz covariance of CKT, see Refs.~\cite{Chen:2012ca,Gao:2017gfq} for the derivation through Wigner functions of relativistic fluids near equilibrium and Refs.~\cite{Mueller:2017arw,Mueller:2017lzw} for a different approach by using the world-line formalism.

Although it is generally believed that the linear response such as CME conductivity $\sigma_B$ is protected by chiral anomaly and independent of the interactions, the nonlinear responses of Weyl fermions could be affected by interactions. Moreover, it has been recently pointed out that the frequency dependent CME conductivity $\sigma_B(\omega)$ could be modified by collisions, the energy shift from the magnetic-moment coupling, and magnetization currents associated with side jumps in the non-equilibrium case \cite{Kharzeev:2009pj,Satow:2014lva,Kharzeev:2016sut}. On the other hand, the coupling independence of the CVE coefficient is in general under debate\footnote{For one of the CVE coefficients only depending on temperature and contributing to axial currents, it is proposed that such a coefficient could be protected by mixed-axial-gravitational anomaly \cite{Landsteiner2011,Landsteiner:2011cp}, while it is found there exists an exceptional case \cite{Hou2012,Golkar:2012kb}. See Refs.\cite{Chowdhury:2015pba,Golkar:2015oxw,Chowdhury:2016cmh} for some following works.}. Similar to the case for background fields, the nonlinear responses involving vorticity may be influenced by interactions as well. In anomalous hydrodynamics, one could classify the possible second-order corrections related to anomalies based on symmetries and thermodynamics \cite{Kharzeev:2011ds}. However, it is useful to utilize microscopic theories such as CKT to obtain such coefficients and analyze their dependence on thermodynamical properties.The second-order nonlinear transport in Weyl-fermion systems have been recently studied with CKT in Refs.~\cite{Gorbar:2016qfh,Chen:2016xtg,Gorbar:2017cwv}. Nonetheless, these studies aim at systems in the absence of collective motion for quasi-particles, which are applicable for Weyl semimetals such that the interaction among quasi-particles is suppressed by their collisions with impurities and phonons. In such cases, the energy-momentum conservation could be violated when neglecting the backreaction upon environments. On the contrary, in QGP, the fluid-like properties of Weyl fermions should be taken due to strong coupling and hydrodynamics impose the energy-momentum conservation. Although in reality the QGP coupling could be too strong for the legitimacy of a kinetic-theory description, there exists a temporal window in early times of heavy ion collisions such that the kinetic theory is applicable to delineate the collective motion of quasi-particles, e.g., see Ref.~\cite{Ebihara:2017suq} for the boost invariant formation of CKT and the so-called chiral circular displacement, and Refs.~\cite{Sun:2016nig,Huang:2017tsq} for the very recent numerical simulations of CKT in heavy ion collisions.

Furthermore, even in Weyl semimetals, the strongly interacting quasi-relativistic plasma could be possibly realized and the hydrodynamics of Weyl fermions should be considered. See e.g., Ref.~\cite{Lucas:2016omy} for such a discussion and the references therein for Dirac fluids in graphene. Therefore, in the comparison with the studies \cite{Gorbar:2016qfh,Chen:2016xtg}, it is also of interest to consider the nonlinear response of CKT under the constraint of hydrodynamics.

In this paper, we investigate second-order nonlinear responses in a chiral fluid with background fields and vorticity by employing the CKT derived from quantum field theories \cite{Hidaka:2016yjf}. Due to the involvement of side-jump terms in collisions, it is nontrivial to show the definition of local equilibrium distributions functions in a proper frame. We tackle this issue first and then focus on nonlinear responses with respect to local fluctuations away from equilibrium led by background fields and local temperature/chemical-potential gradients for right-handed Weyl fermions. In the following, we briefly mention our strategy and summarize some important findings. To solve for the corresponding nonlinear responses perturbatively in gradient expansions, we first apply continuity equations to derive anomalous hydrodynamic equations of motion (EOM) given the first-order transport coefficients obtained from equilibrium Wigner functions. Next, we implement CKT to obtain the non-equilibrium corrections of distributions functions in aid of anomalous hydrodynamic EOM. Given the non-equilibrium distribution functions, we directly compute the charge current and charge density defined by Wigner functions. For simplicity, we neglect viscous corrections and utilize the relaxation-time (RT) approximation for the study of nonlinear responses. Through the paper, we refer \emph{quantum corrections} to the corrections at $\mathcal{O}(\hbar)$ in the Wigner-function approach, which originate from the spin of Weyl fermions and anomalies, 
e.g., some nonlinear responses are from combination of the side-jump and anomalous hydrodynamic
transports. The higher-order corrections in $\mathcal{O}(\hbar^2)$ are beyond the scope of this paper. To give a quick view and simple explanation, here we summarize our findings in short:     
\begin{enumerate}
	\item
	When quantum corrections from side jumps in collisions are considered, the local equilibrium distribution functions can be defined in the frame in accordance to the fluid velocity.
	\item
	Without applying anomalous hydrodynamic EOM, which corresponds to a system breaking energy-momentum conservation due to the interaction with the environment, the quantum corrections of the second-order responses for charge currents under a ``naive'' RT approximation give the terms proportional to ${\bf\nabla\times E}$ and ${\bf E}\times{\bf \nabla\mu}$, which agree with the findings in \cite{Chen:2016xtg,Gorbar:2016qfh}. Here ${\bf E}$ and $\mu$ denote an electric field and a chemical potential.
	\item
    By using anomalous hydrodynamic EOM, which corresponds to an isolated system, the transport coefficients for the ${\bf\nabla\times E}$ and ${\bf E}\times{\bf \nabla\mu}$ terms in charge currents are modified. Furthermore, the ${\bf E}\times\nabla T$ and $(\nabla \mu)\times(\nabla T)$ terms emerge from hydrodynamics, where $T$ denotes temperature.
	\item
	For an inviscid chiral fluid, vorticity does not affect the nonlinear responses in charge currents up to the second order. 
	\item
	Except for the implicit quantum corrections from collisions, the nonlinear quantum correction on charge density vanishes with hydrodynamic EOM in the RT approximation. The result is consistent with the matching condition from the anomalous equation.   
\end{enumerate}
We note that a relevant study was previous presented in Ref.~\cite{Gorbar:2017toh}, whereas the background fields were not included and the subtlety of local equilibrium stemming from side jumps was not discussed therein. 

The paper is organized as following : In Sec.~\ref{Local_Equilibrium}, we investigate the interacting Weyl fermions in local equilibrium from the Wigner-function approach. 
In Sec.~\ref{non_eq_f}, we work out the non-equilibrium distribution functions involving the second-order quantum corrections for an inviscid chiral fluid by using CKT and hydrodynamic EOM. In Sec.~\ref{Nonlinear_Effects}, we implement the non-equilibrium distribution functions to compute the second-order quantum corrections for the charge current and density. We also analyze the corresponding vector/axial currents in high-temperature and large-chemical-potential limits. In the beginning of each section above, we briefly explain our strategy and highlight the key equations and findings, which could be helpful for readers who are not interested in the details of computations. In Sec.~\ref{Discussions_conclusions}, we make brief discussions and outlook. 
For reference, we include the conventions and some widely-used relations in Appendices despite some overlap with the context. We also present the details of some calculations therein. 

Throughout this work, we have choose the metric $\eta_{\mu\nu}=\text{diag}\{+,-,-,-\}$. Therefore, the fluid velocity $u^\mu$ satisfying $u^\mu u_\mu =1$, and the projector is given by $\Theta^{\mu\nu}=\eta^{\mu\nu}-u^{\mu}u^{\nu}$. We also use the Levi-Civita symbol $\epsilon^{\mu\nu\alpha\beta}$ and choose
$\epsilon^{0123}=-\epsilon_{0123}=1$. 

\section{ Local Equilibrium Wigner Functions}\label{Local_Equilibrium}
Before working on nonlinear responses away from equilibrium, first we
would like to review the chiral kinetic theory with side-jump based on the 
Wigner-function approach in Sec. \ref{Wigner}. Because of side jumps, the distribution function
becomes frame dependent stemming from $\hbar$ corrections. Therefore, we need to define the local-equilibrium distribution 
function in a proper way. Consequently, we review the well-defined case
in global equilibrium in Sec. \ref{global}, where the explicit form of the global-equilibrium distribution function for an arbitrary frame is shown in Eq.~(\ref{global_equil_f}). Then we further investigate the case in local equilibrium in Sec. \ref{local}. It turns out that the local-equilibrium distribution function can be introduced in a co-moving frame with the form in Eq.~(\ref{global_equil_f}) such that the collisional kernel in CKT vanishes for at least 2 to 2 scattering. The explicit expression of the corresponding Wigner function in local equilibrium is presented in Eq.~(\ref{L_equil_Wigner}).


\subsection{Wigner Functions and Chiral Kinetic Theory}\label{Wigner}
Wigner functions are defined as the Wigner transformation of lesser/greater propagators,
\begin{eqnarray}
\grave{S}^{<(>)}(q,X)\equiv\int d^4Ye^{\frac{iq\cdot Y}{\hbar}}S^{<(>)}(x,y),
\end{eqnarray}
where $S^<(x,y)=\langle\psi^{\dagger}(y)\psi(x)\rangle$ and $S^>(x,y)=\langle\psi(x)\psi^{\dagger}(y)\rangle$ as the expectation values of fermionic correlators with $Y=x-y$ and $X=(x+y)/2$. Here the gauge link is implicitly embedded to keep gauge invariance and hence $q_{\mu}$ denotes the canonical momentum.
As shown in Ref.~\cite{Hidaka:2016yjf}, by solving Dirac equations up to $\mathcal{O}(\hbar)$, the perturbative solution for the less propagators of right-handed Weyl fermions is given by   
\begin{eqnarray}\label{Wigner_S}
	\grave{S}^{<}(q,X)&=& \bar{\sigma}_{\mu}\grave{S}^{<\mu}(q,X)\notag\\
	&=&\bar{\sigma}_{\mu} 2\pi\bar{\epsilon}(q\cdot n)\left(q^{\mu}\delta(q^2)f^{(n)}_q+\hbar\delta(q^2)S_{(n)}^{\mu\nu}\mathcal{D}_{\nu}f^{(n)}_q
	+\hbar\epsilon^{\mu\nu\alpha\beta}q_{\nu}F_{\alpha\beta}\frac{\partial\delta(q^2)}{2\partial q^2}f^{(n)}_q
	\right),
\end{eqnarray}
where $\bar{\epsilon}(q\cdot n)$ represents the sign of $q\cdot n$, $\bar{\sigma}^{\mu}=(1,-\bm{\sigma})$ are the spin matrices,
and
\begin{eqnarray}
S^{\mu\nu}_{(n)}=\frac{\epsilon^{\mu\nu\alpha\beta}}{2(q\cdot n)}q_{\alpha}n_{\beta} 
\label{S_n_1}
\end{eqnarray}
denotes the spin tensor depending on a frame vector $n^{\mu}$. The choice of a frame corresponds to the choice of an observer and $n^{\mu}$ as a timelike vector represents the four velocity of this observer.
Here we denote $\mathcal{D}_{\beta}f^{(n)}_q=\Delta_{\beta}f^{(n)}_q-\mathcal{C}_{\beta}$, where $\Delta_{\mu}=\partial_{\mu}+F_{\nu\mu}\partial^{\nu}_q$,  $\mathcal{C}_{\beta}=\Sigma_{\beta}^<\bar{f}^{(n)}_q-\Sigma_{\beta}^>f^{(n)}_q$ with $\Sigma_{\beta}^{<(>)}$ being less/greater self-energies and $f^{(n)}_q$ and $\bar{f}^{(n)}_q=1-f^{(n)}_q$ being the distribution functions of incoming and outgoing particles, respectively. In a general case, the distribution functions here are frame dependent, which follows the modified Lorentz transformation between frames\textcolor{blue}{\footnote{The transformation between different frames here is equivalent to the inverse Lorentz transformation of $q^{\mu}$ and $X^{\mu}$. Due to the side-jump term associated with $S_{(n)}^{\mu\nu}$ in Wigner functions, the distribution function is no longer a scalar, which thus undergoes the non-trivial frame transformation or equivalently the modified Lorentz transformation upon phase-space coordinates. One may refer to Ref.~\cite{Hidaka:2016yjf} for more details. As we will discuss later, the explicit expression of distribution functions sometimes may have to be introduced in a particular frame.}}
\begin{eqnarray}
f^{(n')}_q= f^{(n)}_q+\frac{\hbar\epsilon^{\nu\mu\alpha\beta}q_{\alpha}n'_{\beta}n_{\mu}}{2(q\cdot u)(q\cdot n')}\mathcal{D}_{\nu}f^{(n)}_q.
\label{f_n_1}
\end{eqnarray}
Note that both the energy-momentum tensor and currents can be directly obtained from Wigner functions\textcolor{blue}{\footnote{When performing the explicit computations of currents and energy-momentum tensors from Wigner functions, one actually takes normal ordering and drops infinite constants coming from the anti-commutation relation of fermions.}},
\begin{eqnarray}\label{def_J}
T^{\mu\nu}=\int \frac{d^4q}{(2\pi)^4}\big[q^{\mu}\grave{S}^{<\nu}+q^{\nu}\grave{S}^{<\mu}\big],
\quad J^{\mu}=2\int \frac{d^4q}{(2\pi)^4}\grave{S}^{<\mu}.
\end{eqnarray}

In Ref.~\cite{Hidaka:2016yjf}, $n^{\mu}$ is chosen to be independent of $q$ and $X$ except for the part inside collisional kernel, which corresponds to the choice of a global observer in the lab frame.
However, when choosing a local observer, $n^{\mu}$ could depend on spacetime coordinates and the CKT has to be modified.  
Assuming the frame vector $n^{\mu}$ only depends on $X$, we find
\begin{eqnarray}\nonumber
\Delta_{\mu}\grave{S}^{<\mu}
&=&\partial_{\mu}\grave{S}^{<\mu}+F_{\nu\mu}\partial^{\nu}_q\grave{S}^{<\mu}
\\\nonumber
&=&2\pi\bar{\epsilon}(q\cdot n)\Bigg\{\Bigg[
\delta(q^2)q\cdot\Delta  
+
\hbar\delta(q^2)\Big[
F_{\rho \mu}\left(\partial^{\rho}_qS^{\mu\nu}_{(n)}\right)\mathcal{D}_{\nu}
+S^{\mu\nu}_{(n)}(\partial_{\mu}F_{\rho \nu})\partial^{\rho}_{q}
+\left(\partial_{\mu}S^{\mu\nu}_{(n)}\right)\mathcal{D}_{\nu}
\Big]
\\
&&
+\hbar\frac{\partial\delta(q^2)}{\partial q^2}
\Big[2q^{\rho}F_{\mu\rho}S^{\mu\nu}_{(n)}\mathcal{D}_{\nu}
+\frac{1}{2}\epsilon^{\mu\nu\alpha\beta}q_{\nu}F_{\alpha\beta}\Delta_{\mu}\Big]
\Bigg]f^{(n)}_q
-\hbar\delta(q^2)S_{(n)}^{\mu\nu}\Delta_{\mu}\mathcal{C}_{\nu}\Bigg\}
\end{eqnarray}
and 
\begin{eqnarray}\nonumber
\Sigma^<_{\mu}\grave{S}^{>\mu}-\Sigma^>_{\mu}\grave{S}^{<\mu}&=&
2\pi\bar{\epsilon}(q\cdot n)\Bigg(\delta(q^2)q\cdot\mathcal{C}+\hbar\epsilon^{\mu\nu\alpha\beta}\mathcal{C}_{\mu} q_{\nu}F_{\alpha\beta}\frac{\partial\delta(q^2)}{2\partial q^2}
\\
&&+\hbar\delta(q^2)S_{(n)}^{\mu\nu}\big(\Sigma^<_{\mu}\mathcal{D}_{\nu}\bar{f}^{(n)}_q
	-\Sigma^>_{\mu}\mathcal{D}_{\nu}f^{(n)}_q\big)
\Bigg).
\end{eqnarray}
From $\Delta_{\mu}\grave{S}^{<\mu}=\Sigma^<_{\mu}\grave{S}^{>\mu}-\Sigma^>_{\mu}\grave{S}^{<\mu}$,
carrying out similar computations as in Ref.~\cite{Hidaka:2016yjf}, the corresponding CKT takes the form,
\begin{eqnarray}\label{CKT_general}
\delta\left(q^2-\hbar\frac{B\cdot q}{q\cdot n}\right)\Bigg\{\Bigg[
q\cdot\Delta+\hbar\frac{S_{(n)}^{\mu\nu}E_{\mu}}{(q\cdot n)}\mathcal{D}_{\nu}
+\hbar S_{(n)}^{\mu\nu}(\partial_{\mu}F_{\rho \nu})\partial^{\rho}_{q}+\hbar\hat{\Pi}_{(n)}(q,X)\Bigg]f^{(n)}_q
-q\cdot\tilde{\mathcal{C}}\Bigg\}=0,
\end{eqnarray}
where
\begin{eqnarray}
\hat{\Pi}_{(n)}(q,X)=\left(\partial_{\mu}S^{\mu\nu}_{(n)}\right)\mathcal{D}_{\nu}=\frac{\epsilon^{\nu\mu\alpha\beta}}{2q\cdot n}\left(q_{\alpha}(\partial_{\mu}n_{\nu})-\frac{n_{\nu}q_{\alpha}q^{\rho}(\partial_{\mu}n_{\rho})}{q\cdot n}\right)\mathcal{D}_{\beta}
\end{eqnarray}
comes from the choice of a local observer, and

	\begin{eqnarray}
	\tilde{\mathcal{C}}^{\mu}=\mathcal{C}^{\mu}+\hbar\frac{\epsilon^{\mu\nu\alpha\beta}n_{\nu}}{2q\cdot n} 
	\big(\bar{f}^{(n)}_q\Delta^>_{\alpha}\Sigma^{<}_{\beta}-f^{(n)}_q\Delta^<_{\alpha}\Sigma^{>}_{\beta}\big)
	\end{eqnarray}

with $\Delta^{>(<)}_{\mu}=\Delta_{\mu}+\Sigma_{\mu}^{>(<)}$.

Here the electromagnetic fields are defined thorough the frame vector $n^{\mu}$,
\begin{eqnarray}\label{EM_def}
n^{\nu}F_{\mu\nu}=E_{\mu},\quad \frac{1}{2}\epsilon^{\mu\nu\alpha\beta}n_{\nu}F_{\alpha\beta}=B^{\mu},\quad F_{\alpha\beta}=-\epsilon_{\mu\nu\alpha\beta}B^{\mu}n^{\nu}+n_{\beta}E_{\alpha}-n_{\alpha}E_{\beta}.
\end{eqnarray}
Note that $\mathcal{C}_{\mu}$ implicitly incorporates $\hbar$ corrections since it contains at least one internal line of Weyl fermions in the self-energy, which is true for most of realistic scattering processes, and the side-jump term will in general be involved. One can alternatively write the CKT as
\begin{eqnarray}
\Bigg[q\cdot\tilde{\mathcal{D}}+\frac{\hbar S^{\mu\nu}_{(n)}E_{\mu}}{q\cdot n}\mathcal{D}_{\nu}+\hbar S^{\mu\nu}_{(n)}\big(\partial_{\mu}F_{\rho\nu}\big)\partial^{\rho}_q+\hbar\big(\partial_{\mu}S^{\mu\nu}_{(n)}\big)\mathcal{D}_{\nu}
\Bigg]f^{(n)}_q
=0,
\end{eqnarray}
where $\tilde{\mathcal{D}}_{\mu}f^{(n)}_q=\Delta_{\mu} f^{(n)}_q-\tilde{\mathcal{C}_{\mu}}$.
When taking $n^{\mu}=(1,{\bf 0})$ and using the on-shell condition, the CKT reduces to the usual  three-momentum form in Refs.~\cite{Son:2012zy,Hidaka:2016yjf}.

\subsection{Global Equilibrium Cases} \label{global}
It is shown from the semi-classical approach that a global equilibrium distribution function of a rotating Weyl fluid could be defined frame-independently \cite{Chen:2015gta}. We shall first present an equivalent description in Wigner functions and discuss an obstacle for the generalization to local equilibrium.   
Following the definition in Ref.~\cite{Chen:2015gta}, we take the distribution function of right-handed fermions as
\begin{eqnarray}\label{global_equil_f}
	f^{\text{eq}(n)}_q=(e^{g}+1)^{-1},\quad g=\left(\beta q\cdot u- \bar{\mu}+\frac{\hbar S^{\mu\nu}_{(n)}}{2}\partial_{\mu}(\beta u_{\nu})\right),
\end{eqnarray} 
where $\beta=1/T$ is the inverse of temperature $T$ , $\bar{\mu}=\mu/T$ with $\mu$ the charge chemical potential, and $u^{\mu}$ represents the fluid velocity. 
In our further calculations, we also often use the ordinary distribution function without side jumps,
\begin{equation}
f^{(0)}_q=(e^{g_0}+1)^{-1},\;\quad g_0=\beta q\cdot u-\bar{\mu}.
\end{equation}
We shall find that Eq.~(\ref{global_equil_f}) gives rise to the distribution functions in global equilibrium with constant $T$ and $\mu$. For general conditions, we may decompose the derivative of $u_{\nu}$ into symmetric/anti-symmetric parts $\partial_{\mu}u_{\nu}=\sigma_{\mu\nu}+\omega_{\mu\nu}$, where $\sigma_{\mu\nu}=(\partial_{\mu}u_{\nu}+\partial_{\nu}u_{\mu})/2$ and $\omega_{\mu\nu}=(\partial_{\mu}u_{\nu}-\partial_{\nu}u_{\mu})/2$. By introducing the fluid vorticity
\begin{eqnarray}
\omega^{\mu}\equiv\frac{1}{2}\epsilon^{\mu\nu\alpha\beta}u_{\nu}\big(\partial_{\alpha}u_{\beta}\big),
\end{eqnarray}
we may further rewrite the anti-symmetric part as
\begin{eqnarray}
\omega_{\alpha\beta}=-\epsilon_{\alpha\beta\mu\nu}\omega^{\mu}u^{\nu}+\kappa_{\alpha\beta},
\quad
\kappa_{\alpha\beta}=\frac{1}{2}\big(u_{\alpha}u\cdot\partial u_{\beta}-u_{\beta}u\cdot\partial u_{\alpha}\big),
\end{eqnarray}
and the dual tensor

\begin{eqnarray}
\tilde{\omega}^{\mu\nu}=\frac{1}{2}\epsilon^{\mu\nu\alpha\beta}\omega_{\alpha\beta}=\frac{1}{2}\epsilon^{\mu\nu\alpha\beta}u_{\alpha}u\cdot\partial u_{\beta}+\omega^{\mu}u^{\nu}-\omega^{\nu}u^{\mu}.
\end{eqnarray}
By inserting Eq.~(\ref{global_equil_f}) into Eq.~(\ref{Wigner_S}) and using the relation
\begin{eqnarray}\nonumber
	S^{\mu\nu}_{(n)}q^{\rho}\omega_{\nu\rho}+\frac{S^{\rho\nu}_{(n)}}{2}q^{\mu}\omega_{\rho\nu}
	=-\frac{\tilde{\omega}^{\mu\alpha}q_{\alpha}}{2},
\end{eqnarray}
we obtain
\begin{eqnarray}\nonumber\label{Wigner_eq_nframe}
	\grave{S}^{<\mu}&=&2\pi\bar{\epsilon}(q\cdot n)\Bigg\{\Bigg[
	\delta(q^2)\Bigg(q^{\mu}+\frac{\hbar}{2}\big[u^{\mu}(q\cdot\omega)-\omega^{\mu}(q\cdot u)\big]\partial_{q\cdot u}
	-\frac{\hbar}{4}\big(\epsilon^{\mu\nu\alpha\beta}q_{\nu}u_{\alpha}u\cdot\partial u_{\beta}\big)\partial_{q\cdot u}
	\\ \nonumber
	&&
	-\frac{\hbar q^{\mu}S_{(n)}^{\rho\nu}u_{\nu}}{2T^2}(\partial_{\rho}T)
	\partial_{q\cdot u}
	-\hbar S_{(n)}^{\mu\nu}\tilde{E}_{\nu}\partial_{q\cdot u}
	\Bigg)
	+\hbar\epsilon^{\mu\nu\alpha\beta}q_{\nu}F_{\alpha\beta}\frac{\partial\delta(q^2)}{2\partial q^2}
	\Bigg]f^{(0)}_q-\hbar\delta(q^2)S_{(n)}^{\mu\nu}\mathcal{C}_{\nu}\Bigg\},
\\
\end{eqnarray}
where we define
\begin{eqnarray}
\tilde{E}_{\nu}=\mathcal{E}_{\nu}+ \frac{(q\cdot u)}{T}\partial_{\nu}T
-q^{\sigma}(\sigma_{\nu\sigma}+\kappa_{\nu\sigma}),\quad \mathcal{E}_{\nu}=E_{\nu}+T \partial_{\nu}\bar{\mu}.
\end{eqnarray}
For a rotating fluid with constant $T$ and $\mu$ such that $T\partial_{\mu}(\beta u_{\nu})=
-\epsilon_{\mu\nu\alpha\beta}u^{\beta}\omega^{\alpha}$,  the Wigner function reduces to
\begin{eqnarray}\nonumber
\grave{S}^{<\mu}&=&2\pi\bar{\epsilon}(q\cdot n)\Bigg\{\Bigg[
\delta(q^2)\left(q^{\mu}+\frac{\hbar}{2}\big[u^{\mu}(q\cdot\omega)-\omega^{\mu}(q\cdot u)\big]\partial_{q\cdot u}\right)
+\hbar\epsilon^{\mu\nu\alpha\beta}q_{\nu}F_{\alpha\beta}\frac{\partial\delta(q^2)}{2\partial q^2}
\Bigg]f^{(0)}_q
\\
&&
-\hbar\delta(q^2)S_{(n)}^{\mu\nu}\mathcal{C}_{\nu}\Bigg\}.
\end{eqnarray}

The original side-jump term combined with the spin-tensor correction in $f^{\text{eq}}_q$ results in a frame-independent contribution associated with vorticity, which suggests that $f^{(0)}_q$ should be also frame independent and the relevant parameters $T$, $\mu$, and $u^{\mu}$ could be defined universally in arbitrary frames. Given that the collisional kernel vanishes in the center of mass (COM) frame as a no-jump frame \cite{Chen:2015gta,Hidaka:2016yjf}, it should now vanish in an arbitrary frame for global equilibrium. Therefore, the Wigner functions for a purely rotating Weyl fluid in global equilibrium takes the form\footnote{Frame dependence of the sign function $\bar{\epsilon}(q\cdot n)$ does not affect the conclusion.}    
\begin{eqnarray}
\grave{S}^{<\mu}_{\text{geq}}&=&2\pi\bar{\epsilon}(q\cdot n)\Bigg[
\delta(q^2)\left(q^{\mu}+\frac{\hbar}{2}\big[u^{\mu}(q\cdot\omega)-\omega^{\mu}(q\cdot u)\big]\partial_{q\cdot u}\right)
+\hbar\epsilon^{\mu\nu\alpha\beta}q_{\nu}F_{\alpha\beta}\frac{\partial\delta(q^2)}{2\partial q^2}
\Bigg]f^{(0)}_q.
\end{eqnarray}
Nonetheless, when $T$, $\mu$, and $u_{\mu}$ are local parameters, which contribute to not only the vorticity, Eq.~(\ref{Wigner_eq_nframe}) also indicates that these parameters are no longer frame independent under $\hbar$ corrections. Although one can introduce local-equilibrium distribution functions in the COM frame such that the collisional kernel vanishes, which is equivalent to introduce multiple observers for different scattering events with different momenta of incoming and outgoing particles, it is impractical since we may only solve for the distribution function with just one observer in CKT. Technically, in CKT, the Wigner function in $\Delta\cdot\grave{S}^<$ cannot work in the COM frame when $T$, $\mu$, and $u^{\mu}$ depend on the momenta of other scattered particles as a consequence of their $\hbar$ corrections.

Since it is formidable to find a general expression of the local equilibrium distribution function for an arbitrary frame such as Eq.~(\ref{global_equil_f}) in global equilibrium, we may downgrade the problem to seek for the local equilibrium function in a particular frame, from which one can implement the modified Lorentz transformation to write down the corresponding distribution functions in different frames. Fortunately, we find that setting $n^{\mu}=u^{\mu}$ as the co-moving frame with the expression in Eq.~(\ref{global_equil_f}) fits our purpose, which yields the vanishing collisional kernel in 2 to 2 scattering albeit the proof is somewhat technical as we will show in the following subsection.

\subsection{Local Equilibrium Cases}\label{local}
By taking $n^{\mu}=u^{\mu}$ with the distribution function in Eq.~(\ref{global_equil_f}), the Wigner function can be written as
\begin{eqnarray}\label{L_equil_Wigner}\nonumber
	\grave{S}_{\text{leq}}^{<\mu}&=&2\pi\bar{\epsilon}(q\cdot u)\delta(q^2)\Bigg(q^{\mu}
	+\hbar S^{\mu\nu}_{(u)}\Delta_{\nu}
	+\hbar\epsilon^{\mu\nu\alpha\beta}q_{\nu}F_{\alpha\beta}\frac{\partial\delta(q^2)}{2\partial q^2}
	\Bigg)f^{\text{eq}(u)}_q
	\\\nonumber
	&=&2\pi\bar{\epsilon}(q\cdot u)\Bigg[\delta(q^2)\left(q^{\mu}+\frac{\hbar}{2}\big(u^{\mu}(q\cdot\omega)-\omega^{\mu}(q\cdot u)\big)\partial_{q\cdot u}-\hbar S^{\mu\nu}_{(u)}\tilde{E}_{\nu}
	\partial_{q\cdot u}
	\right)
	\\
	&&
	+\frac{\hbar\epsilon^{\mu\nu\alpha\beta}F_{\alpha\beta}}{4}\partial_{q\nu}\delta(q^2)
	\Bigg]f^{(0)}_q.
\end{eqnarray}
Here the collisional corrections in the side-jump term do not contribute to $\grave{S}^{<\mu}_{\text{leq}}$ since $\mathcal{C}_{\mu}$ at $\mathcal{O}(1)$ should be proportional to either $q_{\mu}$ or $u_{\mu}$. 
Given that $f^{(0)}_q=e^{-\beta(q\cdot u-\mu)}\bar{f}^{(0)}_q$, where the bar for $f_q$ here corresponds to the distribution functions for outgoing particles, we can write down a relation between the less and greater propagators,
\begin{eqnarray}\label{KMS}
	\grave{S}_{\text{leq}}^{<\mu}=e^{-\beta(q\cdot u-\mu)}\Bigg(\tilde{S}_{\text{leq}}^{>\mu}-\frac{\hbar}{2T}\Big(u^{\mu}\omega\cdot\grave{S}_{\text{leq}}^{>}-\omega^\mu\grave{S}_{\text{leq}}^{>}\cdot u\Big)
	+\frac{\hbar\epsilon^{\mu\nu\alpha\beta}u_{\nu}}{2T(q\cdot u)}\tilde{E}_{\beta}\grave{S}^>_{\text{leq} \alpha} \Bigg).
\end{eqnarray}
For the leading-order 2 to 2 Coulomb scattering as considered in Ref.~\cite{Hidaka:2016yjf}, using Eq.~(\ref{KMS}), one finds
\begin{eqnarray}\nonumber 
	&&\grave{S}_{\text{leq}}^{>}\cdot\Sigma^<-\grave{S}_{\text{leq}}^{<}\cdot\Sigma^>
	\\
	&&=\hbar\pi\delta(q^2)\int_{q',k,k'}\frac{|\mathcal{M}|^2}{4(k\cdot k')}\bar{f}^{(0)}_q\bar{f}^{(0)}_{q'}f^{(0)}_kf^{(0)}_{k'}
    \Big[\check{\chi}(q,q')+\check{\chi}(q',q)-\check{\chi}(k,k')-\check{\chi}(k',k)\Big], \label{eq:collision}
\end{eqnarray}
where we introduced a compact notation for the integral,
\begin{eqnarray}\nonumber
&&\int_{q',k,k'}=\int \frac{d^4{q'}d^4{k}d^4{k'}}{(2\pi)^{5 }}\delta^{(4)}(q+q'-k-k')\delta(q'^2)\delta(k^2)\delta(k'^2),
\\
&&
\end{eqnarray}
and
\begin{eqnarray}
	\check{\chi}(q,q')=-\frac{1}{T}\Big((q\cdot u)(q'\cdot\omega)-(q\cdot \omega)(q'\cdot u)\Big)+\frac{\epsilon^{\mu\nu\alpha\beta}u_{\nu}}{T(q'\cdot u)}q_{\mu}\tilde{E}_{\beta}(q')q'_{\alpha}.
\end{eqnarray}
Here $|\mathcal{M}|^2$ is the squared matrix element for the 2 to 2 scattering process:
\begin{eqnarray}
|\mathcal{M}|^2=4e^4\left[\frac{(q\cdot q')}{(q\cdot k)}+\frac{(q\cdot q')}{(q\cdot k')}\right]^2.
\end{eqnarray}
The vorticity cancels out in the sum of $\check{\chi}(q,q')$ and $\check{\chi}(q',q)$ appeared in Eq.~(\ref{eq:collision}):
\begin{eqnarray}
	\check{\chi}(q,q')+\check{\chi}(q',q)=\frac{\epsilon^{\mu\nu\alpha\beta}u_{\nu}}{T}q_{\mu}q'_{\alpha}\Bigg(\frac{\tilde{E}_{\beta}(q')}{(q'\cdot u)}-\frac{\tilde{E}_{\beta}(q)}{(q\cdot u)}\Bigg),
\end{eqnarray}
and thus
\begin{eqnarray}\nonumber
	\grave{S}_{\text{leq}}^{>}\cdot\Sigma^<-\grave{S}_{\text{leq}}^{<}\cdot\Sigma^>
	&=&\hbar\pi\delta(q^2)\int_{ q',k,k'}
	\frac{|\mathcal{M}|^2}{4(k\cdot k')}\bar{f}^{(0)}_q\bar{f}^{(0)}_{q'}f^{(0)}_kf^{(0)}_{k'}
	\frac{\epsilon^{\mu\nu\alpha\beta}u_{\nu}}{T}
	\\
	&&\times\Bigg[q_{\mu}q'_{\alpha}\Bigg(\frac{\tilde{E}_{\beta}(q')}{(q'\cdot u)}-\frac{\tilde{E}_{\beta}(q)}{(q\cdot u)}\Bigg)-k_{\mu}k'_{\alpha}\Bigg(\frac{\tilde{E}_{\beta}(k')}{(k'\cdot u)}-\frac{\tilde{E}_{\beta}(k)}{(k\cdot u)}\Bigg)\Bigg].
\end{eqnarray}
It is clear to see that the vorticity-related part, which is actually independent of frames,   vanishes by symmetry. This finding agrees with the case in global equilibrium. Now, we shall deal with the rest part pertinent to $\tilde{E}_{\beta}$.

For convenience, we can work in the local rest frame $u^{\mu}=(1, {\bf u}(X))$ such that ${\bf u}\approx 0$ and $\partial_{\mu}u^0\approx 0$ yet $\partial_{\mu}{\bf u}\neq 0$. Then we find
\begin{eqnarray}\nonumber\label{rest_int}
\grave{S}_{\text{leq}}^{>}\cdot\Sigma^<-\grave{S}_{\text{leq}}^{<}\cdot\Sigma^>
&=&\hbar\pi\delta(q^2)\int_{q',k,k'}
\frac{|\mathcal{M}|^2}{4(k\cdot k')T}\bar{f}^{(0)}_q\bar{f}^{(0)}_{q'}f^{(0)}_kf^{(0)}_{k'}
\\
&&\times
\Bigg[({\bf q\times q'})\cdot\Bigg(\frac{\tilde{\bf E}(q')}{q'_0}-\frac{\tilde{\bf E}(q)}{q_0}\Bigg)-({\bf k\times k'})\cdot\Bigg(\frac{\tilde{\bf E}(k')}{k'_0}-\frac{\tilde{\bf E}(k)}{k_0}\Bigg)\Bigg],
\end{eqnarray}
where $\tilde{\bf E}(q)={\bf E}-T\nabla\bar{\mu}+(\nabla({\bf q\cdot u})+({\bf q\cdot \nabla}){\bf u})/2$. It turns out that this integral actually vanishes, which can be shown based on the symmetry as discussed below. Apparently, the $T\nabla\bar{\mu}$ terms in the integral cancel each other. Now, considering the inversion of spatial momentum and electric fields
(${\bf q}\rightarrow-{\bf q}$, ${\bf q'}\rightarrow-{\bf q'}$, ${\bf k}\rightarrow-{\bf k}$ ,${\bf k'}\rightarrow-{\bf k'},{\bf E}\rightarrow -{\bf E}$). In Eq.~(\ref{rest_int}), we find that the integral is an odd function under the inversion. On the contrary, from the remaining terms in the integral, a non-vanishing collisional kernel should only be proportional to ${\bf q}\cdot\mathcal{\bf E}$ and $({\bf q\cdot \nabla})({\bf q\cdot u})$ and thus should be ``even'' under the inversion.   Accordingly, the integral in Eq.~(\ref{rest_int}) and the full collisional kernel should vanish. We thus conclude that Eq.~(\ref{L_equil_Wigner}) indeed corresponds to the local equilibrium Wigner function at least when considering only 2 to 2 scattering.

\section{Non-Equilibrium Distribution Functions}\label{non_eq_f}
Our final goal is to evaluate second-order quantum corrections on the charge current and density when the system is slightly away from local equilibrium. To handle this problem, we follow the standard strategy: 
expanding all the quantities and evolution equations in the power 
series of $\hbar$ and space-time derivative $\partial$. In the previous
 section, we have defined the distribution function in local equilibrium.
Nevertheless, we have to first derive the non-equilibrium distribution functions led by fluctuations up to the $\mathcal{O}(\hbar\partial^2)$.
 In addition, since we consider a closed system, we shall impose the energy-momentum conservation through anomalous hydrodynamics dictated by continuity equations in Eq.~(\ref{conti_eq}) as constrains for CKT. 
 
 In Sec. \ref{Anomalous_Hydrodynamics}, we derive the equations of motion (EOM)
 for anomalous hydrodynamics necessary for the study of $\hbar$ corrections on non-equilibrium distribution functions.
 Those anomalous-hydrodynamic EOM will govern the dynamics of free thermodynamic parameters such as $T$, $\mu$, and $u^{\mu}$ in local-equilibrium distribution functions as shown in Eq.~(\ref{D0T_D0mu}). 
 Later, these relations have to be applied when solving the non-equilibrium distribution function perturbtively from CKT.
	
Subsequently, in Sec. \ref{Non_equ_Responses} the non-equilibrium distribution function is perturbatively solved from CKT by using an ansatz in Eq.~(\ref{f_ansatz}) with the RT approximation. The corresponding solutions are presented in Eqs.~(\ref{delfC}) and (\ref{delfQ}), where the quantum part is further composed of three pieces coming from the $\hbar$ corrections in CKT, hydrodynamic EOM, and postulated terms in collisions, respectively.
\subsection{ Anomalous Hydrodynamic Equations}\label{Anomalous_Hydrodynamics}
Now, armed with the local equilibrium Wigner function in Eq.~(\ref{L_equil_Wigner}), we may first proceed to reproduce first-order anomalous transport coefficients in hydrodynamics, which have been studied from various approaches (e.g., see Refs.~\cite{Gao:2012ix, Landsteiner:2012kd} and the references therein). On the other hand, we will also derive the hydrodynamic EOM with quantum corrections, which further contribute to the second-order transport.  

In local equilibrium, the constitutive relations for energy-momentum tensors and charge currents are given by
\begin{eqnarray}
T^{\mu\nu}=u^{\mu}u^{\nu}\epsilon-p\Theta^{\mu\nu}+\Pi_\text{non}^{\mu\nu}+\Pi^{\mu\nu}_\text{dis},\quad J^{\mu}=N_0u^{\mu}+v_\text{non}^{\mu}+v_\text{dis}^{\mu},
\end{eqnarray}
where $\Theta^{\mu\nu}=\eta^{\mu\nu}-u^{\mu}u^{\nu}$ and $\epsilon$ and $p$ denote the energy density and pressure, respectively. Here the subindices ``non'' and ``dis'' represent the non-dissipative and dissipative corrections, respectively. 
The non-dissipative corrections come from anomalous transport, which can be written as 
\begin{eqnarray}
\Pi_\text{non}^{\mu\nu}=\hbar\xi_{\omega}\big(\omega^{\mu}u^{\nu}+\omega^{\nu}u^{\mu}\big)
+\hbar\xi_{B}\big(B^{\mu}u^{\nu}+B^{\nu}u^{\mu}\big),
\quad v_\text{non}^{\mu}=\hbar\sigma_B B^{\mu}+\hbar\sigma_{\omega}\omega^{\mu},
\end{eqnarray}
where $\xi_{\omega}$ and $\xi_{B}$ contribute to the heat conductivity and $\sigma_B$ and $\sigma_{\omega}$ corresponds to the charge conductivity of CME and CVE, respectively. Note that the electromagnetic fields are defined in Eq.~(\ref{EM_def}). 
Such decompositions we applied are more convenient to be embedded into CKT, which are distinct from the Landau frame implemented in the previous studies of anomalous hydrodynamics such as in Ref.~\cite{Son:2009tf,Yamamoto:2015ria}.

We may compute the non-dissipative contributions of energy-momentum tensors and currents from the Wigner functions in local equilibrium,
\begin{eqnarray}\nonumber
T_{\text{leq}}^{\mu\nu}&=&u^{\mu}u^{\nu}\epsilon-p\Theta^{\mu\nu}+\Pi_\text{non}^{\mu\nu}=\int \frac{d^4q}{(2\pi)^4}\big(q^{\mu}\grave{S}^{<\nu}_{\text{leq}}+q^{\nu}\grave{S}^{<\mu}_{\text{leq}}\big),
\\
J_{\text{leq}}^{\mu}&=&N_0u^{\mu}+v_\text{non}^{\mu}=2\int \frac{d^4q}{(2\pi)^4}\grave{S}_{\text{leq}}^{<\mu},
\end{eqnarray}
whereas the dissipative parts stem from non-equilibrium corrections associated with collisions.
In practice, it is more convenient to work in the local rest frame to derive the transport coefficients and plug them back into the constitutive relations. By employing Eq.~(\ref{L_equil_Wigner}) and carrying out direct computations, we find 
\begin{eqnarray}\label{coefficients1}
\epsilon=3p=T^4\Big(\frac{7\pi^2}{120}+\frac{\bar{\mu}^2}{4}+\frac{\bar{\mu}^4}{8\pi^2}\Big),
\quad N_0=\frac{T^3}{6}\left(\bar{\mu}+\frac{\bar{\mu}^3}{\pi^2}\right),
\end{eqnarray}
\begin{eqnarray}
\sigma_{\omega}=\frac{T^2}{12}\Bigg(1+\frac{3\bar{\mu}^2}{\pi^2}\Bigg)
,\quad \sigma_B=\frac{\mu}{4\pi^2},
\end{eqnarray}
and
\begin{eqnarray}
\xi_{\omega}=\frac{T^3}{6}\Big(\bar{\mu}+\frac{\bar{\mu}^3}{\pi^2}\Big)=N_0,\quad \xi_B=\frac{ T^2}{24}\Big(1+\frac{3\bar{\mu}^2}{\pi^2}\Big)=\frac{\sigma_{\omega}}{2}.
\end{eqnarray}
Note that $\tilde{E}^{\mu}$ term in $\grave{S}^{<\mu}_{\text{leq}}$ does not contribute to $v^{\mu}_\text{non}$ and $\Pi^{\mu\nu}_\text{non}$, which can be shown in direct calculations. The anomalous coefficients obtained above agree with what have been found previously e.g., from Kubo formulae \cite{Landsteiner:2012kd} or hydrodynamics with second-law of thermodynamics \cite{Neiman:2010zi}.

Subsequently, using the continuity equations,
\begin{eqnarray}\label{conti_eq}
\partial_{\mu}T^{\mu\nu}=F^{\nu\rho}J_{\rho},\quad \partial_{\mu}J^{\mu}=\frac{\hbar}{4\pi^2}({\bf E\cdot B}),
\end{eqnarray}
and taking the projections, we obtain
\begin{eqnarray}\nonumber \label{de0}
u_{\nu}\partial_{\mu}T^{\mu\nu}&=&(\epsilon+p)\partial\cdot u+u\cdot\partial \epsilon
+\hbar\big(\omega\cdot\partial\xi_{\omega}+\xi_{\omega}\partial\cdot\omega
+(\omega\cdot u)(u\cdot\partial)\xi_{\omega}+\xi_{\omega}(\omega\cdot u)(\partial\cdot u)
\\\nonumber
&&
+\xi_{\omega}u_{\nu}u\cdot\partial\omega^{\nu}
\big)
+(\omega\leftrightarrow B)+u_{\nu}\partial_{\mu}\Pi^{\mu\nu}_\text{dis}
\\\nonumber
&=&(\epsilon+p)\partial\cdot u+u\cdot\partial \epsilon
+\hbar\big(\omega\cdot\partial\xi_{\omega}+\xi_{\omega}\partial\cdot\omega
+\xi_{\omega}u_{\nu}u\cdot\partial\omega^{\nu}
\big)+(\omega\leftrightarrow B)+u_{\nu}\partial_{\mu}\Pi^{\mu\nu}_\text{dis}
\\
&=&u_{\nu}F^{\nu\rho}J_{\rho},
\end{eqnarray}
and
\begin{eqnarray}\nonumber
\Theta^{\alpha}_{\mbox{ }\nu}\partial_{\mu}T^{\mu\nu}&=&(\epsilon+p)u\cdot\partial u^{\alpha}
-\partial^{\alpha}p+u^{\alpha}(u\cdot\partial p)+
\hbar\big[\xi_{\omega}(\omega\cdot\partial)u^{\alpha}+\xi_{\omega}\omega^{\alpha}\partial\cdot u+\omega^{\alpha}(u\cdot\partial)\xi_{\omega}
\\\nonumber
&&+\xi_{\omega}(u\cdot\partial)\omega^{\alpha}-u^{\alpha}\xi_{\omega}u_{\nu}u\cdot\partial\omega^{\nu}
+(\omega\leftrightarrow B)
\big]+\Theta^{\alpha}_{\mbox{ }\nu}\partial_{\mu}\Pi^{\mu\nu}_\text{dis}
\\
&=&(F^{\alpha\rho}+u^{\alpha}E^{\rho})J_{\rho}. \label{delta_TEM}
\end{eqnarray}
For convenience, we will work in the local rest frame and our goal is to solve for the time derivatives of parameters $T$, $\bar{\mu}$, and ${\bf u}$. In addition, the incorporation of 
$v^{\mu}_\text{dis}$ and $\Pi^{\mu\nu}_\text{dis}$ 
will result in $\mathcal{O}(\partial^2)$ corrections on $\partial_0{\bf u}$, $\partial_0T$, and $\partial_0\bar{\mu}$. Consequently, 
$v^{\mu}_\text{dis}$ and $\Pi^{\mu\nu}_\text{vis}$, 
which could yield quantum corrections at least for $\mathcal{O}(\hbar\partial^3)$ in CKT, will be omitted.
The calculations in a covariant form are shown in 
Appendix~\ref{Ahydro}. For simplicity, we further drop the viscous corrections.
Working in the local rest frame and implementing the constitutive relations, the continuity equations give rise to
\begin{eqnarray} \label{DN0}
\partial_0 N_0
=-\hbar\big[{\bm\omega}\cdot\nabla \sigma_{\omega}+2\sigma_{\omega}{\bm \omega}\cdot \partial_0{\bf u}+{\bf B\cdot\nabla}\sigma_{B}+\sigma_{B}\partial\cdot B\big]+\frac{\hbar}{4\pi^2}({\bf E\cdot B}),
\end{eqnarray}
\begin{eqnarray}\nonumber
\partial_0\epsilon
&=&-\hbar\big({\bm\omega\cdot\nabla}\xi_{\omega}+\xi_{\omega}\partial\cdot \omega
+\xi_{\omega}{\bm\omega}\cdot\partial_0{\bf u}\big)
-\hbar\big({\bf B\cdot\nabla}\xi_{B}+\xi_{B}\partial\cdot B
+\xi_{B}{\bf B}\cdot\partial_0{\bf u}\big)
\\
&&+\hbar \big(\sigma_{\omega}({\bf E\cdot\bm{\omega}})+\sigma_B({\bf E\cdot B})\big),
\end{eqnarray}
\begin{eqnarray} \label
-(\epsilon+p)\partial_0{\bf u}
&=&\nabla p-N_0{\bf E}
+\hbar\big[{\bm\omega}\partial_0\xi_{\omega}+\xi_{\omega}\partial_0{\bm\omega}
+{\bf B}\partial_0\xi_{B}+\xi_{B}\partial_0{\bf B}
-\big(\xi_B+\sigma_{\omega}\big)({\bm{\omega}\times\bf B})
\big].
\end{eqnarray}
In aid of Bianchi identity $\partial_{\mu}\tilde{F}^{\mu\nu}=0$ and $\partial_{\mu}\tilde{\omega}^{\mu\nu}=0$, which takes explicit forms as
\begin{eqnarray}\label{Bianchi_identity_EM}
\partial\cdot B-2{\bf E\cdot\bm\omega}-{\bf B}\cdot\partial_0{\bf u}=0,\quad
\partial_0{\bf B}+{\bf B\times\bm{\omega}}+({\bf \nabla\times E}-{\bf E}\times\partial_0{\bf u})=0,
\end{eqnarray}
and
\begin{eqnarray}\label{Bianchi_identity_omega}
\partial\cdot\omega-2{\bm{\omega}}\cdot\partial_0{\bf u}=0,\quad
\partial_0{\bm\omega}-\frac{1}{2}\nabla\times(\partial_0{\bf u})=0,
\end{eqnarray}
we perturbatively solve the continuity equations up to $\mathcal{O}(\hbar)$ and obtain the hydrodynamic EOM,
	\begin{eqnarray}\label{D0T_D0mu}
	\frac{\partial_0T}{T}=\hbar\bm{\mathcal{E}}\cdot\Big(\tilde{T}_B{\bf B}+\tilde{T}_{\omega}\bm{\omega}T\Big),\quad
	\partial_0\bar{\mu}=\hbar\bm{\mathcal{E}}\cdot\Big(\tilde{\mu}_B{\bf B}+\tilde{\mu}_{\omega}\bm{\omega}T\Big),
	\end{eqnarray}
	and
	\begin{eqnarray}\label{D0ui_full}\nonumber
	&&\partial_0{\bf u}=\partial_0{\bf u}^{(0)}+\hbar\partial_0\delta {\bf u},
	\quad
	\partial_0{\bf u}^{(0)}=-\frac{\nabla T}{T}+\frac{N_0\bm{\mathcal{E}}}{4p},
	\\
	&&\hbar\partial_0\delta {\bf u}=\hbar\Bigg(\tilde{U}_E\nabla\times{\bf E}+\tilde{U}_T\frac{{\bf E}\times\nabla T}{T}+\tilde{U}_{\bar{\mu}}{\bf E}\times\nabla\bar{\mu}\Bigg),
	\end{eqnarray}
where the coefficients involved have the following dimensions in energy, $\tilde{T}_{B(\omega)}:\mathcal{O}(T^{-3})$, $\tilde{\mu}_{B(\omega)}:\mathcal{O}(T^{-3})$, $\tilde{U}_{E,T,\bar{\mu}}:\mathcal{O}(T^{-2})$.
The explicit forms of these coefficients read
\begin{eqnarray}\nonumber 
\tilde{U}_E&=&\tilde{U}_T=\frac{T^2}{96 p\pi^2} 
\Bigg[(3 \bar{\mu}^2+\pi^2)-\frac{ N_0\bar{\mu}T(\bar{\mu}^2+ \pi ^2)}{2p}\Bigg], 
\\
\tilde{U}_{\bar{\mu}}&=&\frac{T^3}{384p^2\pi^2}\Bigg[N_0(3 \bar{\mu}^2+\pi^2)+
 \bar{\mu}T(\bar{\mu}^2+ \pi ^2)\Big(2\sigma_{\omega}-\frac{N_0^2}{p}\Big)\Bigg], \label{UE_UU}
\end{eqnarray}

\begin{eqnarray}\nonumber 
\tilde{T}_B&=&-\frac{15\bar{\mu}\Big(15 \bar{\mu}^4+50 \bar{\mu}^2 \pi ^2+19 \pi ^4\Big)}{2 \left(15 \bar{\mu}^4+6 \bar{\mu}^2 \pi ^2+7 \pi ^4\right) \left(15 \bar{\mu}^4+30 \bar{\mu}^2 \pi ^2+7 \pi ^4\right) T^3},
\\
\tilde{T}_{\omega}&=&-\frac{5\Big(45 \bar{\mu}^6+75 \bar{\mu}^4 \pi ^2-9 \bar{\mu}^2 \pi ^4-7 \pi ^6\Big)}{\left(15 \bar{\mu}^4+6 \bar{\mu}^2 \pi ^2+7 \pi ^4\right) \left(15 \bar{\mu}^4+30 \bar{\mu}^2 \pi ^2+7 \pi ^4\right) T^3}, \label{TB_TO}
\end{eqnarray}

\begin{eqnarray}\nonumber 
\tilde{\mu}_B&=&\frac{3 \Big(75 \bar{\mu}^6+375 \bar{\mu}^4 \pi ^2+285 \bar{\mu}^2 \pi ^4+49 \pi ^6\Big)}{2 \left(15 \bar{\mu}^4+6 \bar{\mu}^2 \pi ^2+7 \pi ^4\right) \left(15 \bar{\mu}^4+30 \bar{\mu}^2 \pi ^2+7 \pi ^4\right) T^3},
\\
\tilde{\mu}_{\omega}&=&\frac{3 \Big(75 \bar{\mu}^6+225 \bar{\mu}^4 \pi ^2+65 \bar{\mu}^2 \pi ^4-21 \pi ^6\Big)}{\left(15 \bar{\mu}^4+6 \bar{\mu}^2 \pi ^2+7 \pi ^4\right) \left(15 \bar{\mu}^4+30 \bar{\mu}^2 \pi ^2+7 \pi ^4\right) T^3}. \label{UB_UO}
\end{eqnarray}
On can make a quick crosscheck with what have been found in the Landau frame, for which we shall shift the fluid velocity as
\begin{eqnarray}
\tilde{u}^{\mu}=u^{\mu}+\hbar\frac{\xi_{\omega}\omega^{\mu}+\xi_{B}B^{\mu}}{\epsilon+p},
\end{eqnarray}
which yields
\begin{eqnarray}
J_{\text{leq}}^{\mu}=N_0\tilde{u}^{\mu}+\hbar\tilde{\sigma}_{\omega}\omega^{\mu}+\hbar\tilde{\sigma}_{B}B^{\mu},
\end{eqnarray}
where
\begin{eqnarray}
\tilde{\sigma}_{\omega}=\sigma_{\omega}-\frac{N_0\xi_{\omega}}{\epsilon+p},\quad 
\tilde{\sigma}_{B}=\sigma_{B}-\frac{N_0\xi_{B}}{\epsilon+p}.
\end{eqnarray}
By using the result in Eq~.(\ref{D0ui_full}), we derive
	\begin{eqnarray}
	\partial_0\tilde{\bf u}=-\frac{\nabla T}{T}+\frac{N_0\bm{\mathcal{E}}}{4p}-\frac{\hbar \sigma_{\omega}}{8 p}
	\bm{\omega}\times {\bf B},
	\end{eqnarray}
which is in accordance with the finding in Ref.~\cite{Yamamoto:2015ria}. As discussed therein, this equation is responsible for the Chiral Alfven waves.

\subsection{ Non-equilibrium Responses from CKT}\label{Non_equ_Responses}
In this section, we would like to investigate the second-order corrections in terms of gradient expansions on the distribution functions when a system is slightly driven away from local equilibrium by external background fields and temperature/chemical-potential gradients. We work in the $n^{\mu}=u^{\mu}$ frame and the CKT in Eq.~(\ref{CKT_general}) can be written as
\begin{eqnarray}
\Bigg[
q\cdot\Delta+\hbar\frac{S_{(u)}^{\mu\nu}E_{\mu}}{(q\cdot u)}\Delta_{\nu}
+\hbar S_{(u)}^{\mu\nu}(\partial_{\mu}F_{\rho \nu})\partial^{\rho}_{q}+\hbar\big(\partial_{\rho}S^{\rho\mu}_{(u)}\big)\Delta_{\mu}\Bigg]f^{(u)}_q
=\mathcal{C}_\text{full},
\end{eqnarray}
where
\begin{eqnarray}
\mathcal{C}_\text{full}=q^{\mu}\tilde{\mathcal{C}}_{\mu}+\hbar\frac{S_{(u)}^{\mu\nu}E_{\mu}}{(q\cdot u)}\mathcal{C}_{\nu}+\hbar\big(\partial_{\rho}S^{\rho\mu}_{(u)}\big)\mathcal{C}_{\mu}.
\end{eqnarray}
The explicit form of $\partial_{\rho}S^{\rho\mu}_{(u)}$ is evaluated as
\begin{eqnarray}
\partial_{\rho}S^{\rho\mu}_{(u)}=\frac{1}{2}\Bigg[\omega^{\mu}-\frac{(q\cdot\omega)u^{\mu}}{q\cdot u}
-\frac{(q\cdot\omega)q^{\mu}}{(q\cdot u)^2}\Bigg]+
\frac{\epsilon^{\mu\nu\alpha\beta}}{2q\cdot u}\Bigg(q_{\alpha}\kappa_{\beta\nu}
-\frac{u_{\nu}q_{\alpha}q^{\rho}}{q\cdot u}(\sigma_{\beta\rho}+\kappa_{\beta\rho})\Bigg).
\end{eqnarray}
We then introduce the perturbation on distribution functions, 
\begin{eqnarray}\label{f_ansatz}
f^{(u)}_q-f^\text{eq}_q=\delta f_q=\delta f^{(c)}_{q}+\hbar\delta f^{(Q)}_{q},
\end{eqnarray}
in which we further decompose the classical and quantum corrections. 
In order to just analyze non-equilibrium transport qualitatively and derive analytic expressions, we will implement the RT approximation to simplify the collisional terms. Nevertheless, in non-equilibrium states, the nonzero collisional terms also include $\hbar$ corrections. For completeness, we approximate
\begin{eqnarray}\label{RT_approx}
\mathcal{C}_\text{full}=-\frac{1}{\tau_R}\Big(q\cdot u+\hbar \frac{q^{\mu}\mathcal{A}_{\mu}}{(q\cdot u)^2}\Big)\delta f_q,
\end{eqnarray}
where $\mathcal{A}_{\mu}$ could be proportional to $B_{\mu}$, $\omega_{\mu}$, $E_{\mu}$, etc. at $\mathcal{O}(\partial)$. Here the relaxation time $\tau_R$ is regarded as a dimensionful constant under the ``naive'' RT approximation, which describes how long the system
could return to the equilibrium state.
The perturbative solution then takes the form,
\begin{eqnarray}\label{pert_fq}
	\delta f_q=-\frac{\tau_R}{q\cdot u}\left(1-\frac{\hbar q\cdot\mathcal{A}}{q\cdot u}\right)\Bigg[
	q\cdot\Delta+\hbar\frac{S_{(u)}^{\mu\nu}E_{\mu}}{(q\cdot u)}\Delta_{\nu}
	+\hbar S_{(u)}^{\mu\nu}(\partial_{\mu}F_{\rho \nu})\partial^{\rho}_{q}+\hbar\big(\partial_{\rho}S^{\rho\mu}_{(u)}\big)\Delta_{\mu}\Bigg]f^\text{eq}_q.
\end{eqnarray}
In general, the classical part of the non-equilibrium distribution function $\delta f^{(c)}$ should also involve second-order terms responsible for e.g., classical Hall effects. These terms may contribute to higher-order quantum transport starting from $\mathcal{O}(\hbar\partial^3)$ in currents, while we only focus on second-order quantum corrections and hence omit such terms. Consequently, we have
\begin{eqnarray}\label{delfC}
\delta f^{(c)}_q=\frac{\tau_R q\cdot\tilde{E}}{q\cdot u}\partial_{q\cdot u}f^{(0)}_q.
\end{eqnarray}

Based on Eq.~(\ref{pert_fq}), one finds that the quantum corrections in the non-equilibrium distribution function can be decomposed into three parts,
$\hbar\delta f^{(Q)}_q=\delta f^{\mathcal{K}}_q+\delta f^{\mathcal{H}}_q+\delta f^{\mathcal{C}}_q$, where 
\begin{eqnarray}\nonumber\label{delfQ}
&&\delta f^{\mathcal{K}}_q=-\frac{\tau_R}{q\cdot u}\Bigg\{\Bigg[
q\cdot\Delta+\hbar\frac{S_{(u)}^{\mu\nu}E_{\mu}}{(q\cdot u)}\Delta_{\nu}
+\hbar S_{(u)}^{\mu\nu}(\partial_{\mu}F_{\rho \nu})\partial^{\rho}_{q}+\hbar\big(\partial_{\rho}S^{\rho\mu}_{(u)}\big)\Delta_{\mu}\Bigg]f^\text{eq}_q-q\cdot\Delta f^{(0)}_q
\Bigg\},
\\\nonumber
&&\delta f^{\mathcal{H}}_q=\tau_R\Bigg[Tu\cdot\partial\bar{\mu}+\frac{(q\cdot u) u\cdot \partial T}{T}-\hbar q^{\mu}u\cdot \partial\delta u_{\mu}\Bigg]\partial_{q\cdot u}f^{(0)}_q,
\\
&&\delta f^{\mathcal{C}}_q=-\frac{\hbar\tau_R}{(q\cdot u)^4}(q\cdot\tilde{E})(q\cdot\mathcal{A})\partial_{q\cdot u}f^{(0)}_q.
\end{eqnarray}
Here, $\delta f^{\mathcal{K}}_q$ and $\delta f^{\mathcal{H}}_q$ come from the explicit $\hbar$ corrections in CKT and the $\hbar$ corrections from hydrodynamic EOM, respectively. However, the last one $\delta f^{\mathcal{C}}_q$ stems from the $\hbar$ corrections of collisions. Note that in the local rest frame, $q\cdot\mathcal{A}$ must be linear to ${\bf q}$ given that it is at $\mathcal{O}(\partial)$. We may accordingly take $q\cdot\mathcal{A}=-{\bf q}\cdot\bm{\mathcal{A}}$ with $\bm{\mathcal{A}}=\mathcal{A}^i$ without the loss of generality. 

Now, $\delta f^{\mathcal{H}}_q$ can be read out from hydrodynamic EOM in Eqs.~(\ref{D0T_D0mu}) and (\ref{D0ui_full}). In the local rest frame, we have 
\begin{eqnarray}\nonumber
\delta f^{\mathcal{H}}_q&=&\hbar\tau_R\Bigg[\bm{\mathcal{E}}\cdot\Big(\big(T\tilde{\mu}_B+q_0\tilde{T}_B\big){\bf B}
+\big(T\tilde{\mu}_{\omega}+q_0\tilde{T}_{\omega}\big)\bm{\omega}T\Big)
\\
&&
+{\bf q}\cdot\Bigg(\tilde{U}_E{\bf\nabla \times E}+\tilde{U}_T\frac{{\bf E}\times\nabla T}{T}+\tilde{U}_{\bar{\mu}}{\bf E\times\nabla}\bar{\mu}\Bigg)\Bigg]\partial_{q_0}f^{(0)}_q.
\end{eqnarray}
The derivation of $\delta f^{\mathcal{K}}_q$ is somewhat complicated yet straightforward, for which we present some details of calculations in Appendix~\ref{Ahydro}. By using Eq.~(\ref{2nd_order_cal}), in the local rest frame, we find 
\begin{eqnarray}\nonumber\label{del_fk}
\delta f^{\mathcal{K}}_q&=&-\frac{\hbar\tau_R}{2q_0}
\Bigg\{\Bigg[\bm{\omega}\cdot\bm{\mathcal{E}}-\frac{q_0\bm{\omega}\cdot\nabla T}{T}
+\frac{({\bf q}\cdot\bm{\omega})({\bf q}\cdot\bm{\mathcal{E}})}{q_0^2}
+\frac{(q\cdot\partial)(q\cdot\omega)}{q_0}
-\frac{({\bf q}\cdot\bm{\omega})}{q_0}\Big({\bf q}\cdot\bm{\mathcal{E}}-\frac{q_0{\bf q\cdot\nabla}T}{T}\Big)\partial_{q0}
\\\nonumber
&&
+\left(\frac{{\bf q}}{q_0}\right)\cdot\left({\bf E}\times\left(\frac{T\nabla\bar{\mu}}{q_0}+\frac{\nabla T}{T}\right)+\nabla\times{\bf E}\right)\Bigg]
\\\nonumber
&&
-\Bigg[\frac{({\bf q\cdot\bm\omega})}{q_0}\Bigg(\Big(T\partial_0\bar{\mu}+{\bf q}\cdot\partial_0 {\bf u}\Big)(2-q_0\partial_{q0})-\frac{q_0\partial_0T}{T}(1-q_0\partial_{q0})\Bigg)
+q_0\bm{\omega}\cdot\partial_0{\bf u}
\\
&&
+\frac{(\partial_0{\bf u})\times {\bf q}}{q_0}\cdot \left(\bm{\mathcal{E}}-\frac{q_0\nabla T}{T}\right)
\Bigg]
\Bigg\}\partial_{q0}f^{(0)}_q.
\end{eqnarray}
Here the terms in the first two lines of Eq.~(\ref{del_fk}) are independent of hydrodynamic EOM. When including the hydrodynamic corrections in Eq.~(\ref{del_fk}), only the leading-order hydrodynamic EOM are needed. Therefore, one can in fact drop the terms containing $\partial_0T$ and $\partial_0\bar{\mu}$ in Eq.~(\ref{del_fk}), which are at the order of $\mathcal{O}(\hbar)$ shown in Eq.~(\ref{D0T_D0mu}) while we keep them here simply for completeness.  

\section{ Nonlinear Effects in Non-equilibrium Currents}\label{Nonlinear_Effects}
After deriving the second-order non-equilibrium distribution function from the previous section, we will then insert it into the Wigner function and evaluate the second-order quantum corrections upon the charge current and density in Sec. \ref{charge} and Sec. \ref{charge density}, respectively. The final result of anomalous Hall currents triggered by electric fields and temperature/chemical-potential gradients is shown in Eq.~(\ref{charge_J_hydro}). On the other hand, we also show that the non-equilibrium correction on charge density can only be contributed by the postulated $\hbar$ corrections in collisions as shown in Eq.~(\ref{J0Q}), which agrees with the matching condition led by Eq.~(\ref{matching_cond}) from CKT. Finally, in Sec.~\ref{va_currents}, with the inclusion of left-handed fermions, we present the vector/axial charge currents with second-order quantum corrections in high/low-temperature limits.
\subsection{Charge Currents} \label{charge}
Given the non-equilibrium distribution function, we may start to compute the second-order responses in charge currents. We shall focus on just quantum corrections and neglect classical effects. From Eqs.~(\ref{Wigner_S}) and (\ref{def_J}), the quantum corrections of the non-equilibrium current read 
\begin{eqnarray}
\delta J_Q^{\mu}=2\hbar\int\frac{d^4q}{(2\pi)^3}\bar{\epsilon}(q\cdot u)\delta(q^2)\Bigg[ q^{\mu}\delta f^{(Q)}_q
-\frac{1}{2}\Bigg(\epsilon^{\mu\nu\alpha\beta}\frac{u_{\nu}q_{\alpha}}{q\cdot u}\Delta_{\beta}
+\epsilon^{\mu\nu\alpha\beta}F_{\alpha\beta}\frac{\partial_{q\nu}}{2}\Bigg)
\delta f^{(c)}_q
\Bigg].
\end{eqnarray}
In the local rest frame, the charge density and charge current accordingly take the form,
\begin{eqnarray}\nonumber\label{delJQl}
\delta J_Q^{0}&=&2\hbar\int\frac{d^4q}{(2\pi)^3}\bar{\epsilon}(q_0)\delta(q^2)\left(q_0\delta f^{(Q)}_q+\frac{\bf B}{2}\cdot\nabla_{\bf q}\delta f^{(c)}_q\right),
\\\nonumber
\delta {\bf J}_Q&=&2\hbar\int\frac{d^4q}{(2\pi)^3}\bar{\epsilon}(q_0)\delta(q^2)\Bigg[{\bf q}\delta f^{(Q)}_q
-\frac{1}{2q_0}\left({\bf q}\times\nabla+({\bf q\times E})\partial_{q0}-q_0{\bf E}\times\nabla_{\bf q}\right)\delta f^{(c)}_q
\\
&&+\frac{1}{2q_0}\left(q_0{\bf B}\partial_{q0}
-({\bf q\cdot B})\nabla_{\bf q}+{\bf B}({\bf q}\cdot\nabla_{\bf q})
\right)
\delta f^{(c)}_q
\Bigg],
\end{eqnarray} 
where $\nabla_{\bf q}=\partial/\partial {\bf q}$. We will now evaluate the charge current in the following. 

We shall first consider the current led by $\delta f^{(Q)}_q$, which consists of three parts shown in Eq.~(\ref{delfQ}). We first compute the contribution from $\delta f^{\mathcal{K}}_q$, which yields
\begin{eqnarray}\nonumber
\delta {\bf J}_{\mathcal{K}}&=&2\int\frac{d^4q}{(2\pi)^3}\bar{\epsilon}(q_0)\delta(q^2){\bf q}\delta f^{\mathcal{K}}_q
\\\nonumber
&=&-\frac{\hbar\tau_R}{3}\int\frac{d^4q}{(2\pi)^3}\bar{\epsilon}(q_0)\delta(q^2)
\Bigg[\left({\bf\nabla\times E}+{\bf E}\times\left(\frac{T\nabla\bar{\mu}}{q_0}+\frac{\nabla T}{T}\right)\right)
-q_0\partial_0\bm{\omega}
\\\nonumber
&&
-\left({\bf E}-T\nabla\bar{\mu}-\frac{q_0\nabla T}{T}\right)\partial_0{\bf u}
-\bm{\omega}\Bigg(T\partial_0\bar{\mu}(2-q_0\partial_{q0})+\frac{q_0\partial_0T}{T}(1-q_0\partial_{q0})\Bigg)
\Bigg]\partial_{q0}f^{(0)}_q.
\\
\end{eqnarray}
Implementing the useful integrals in Appendix~\ref{convention} and the Bianchi identity in Eq.~(\ref{Bianchi_identity_omega}), we find
\begin{eqnarray}\nonumber
\delta {\bf J}_{\mathcal{K}}
&=&-
\frac{\tau_R\hbar}{12\pi^2}\Bigg[
-\mu\nabla\times{\bf E}
-{\bf E}\times\Big(T\nabla\bar{\mu}+\bar{\mu}\nabla T-\mu\partial_0{\bf u}\Big)
-T(\mu\nabla\bar{\mu}+I_1\nabla T)\partial_0{\bf u}
\\\nonumber
&&
+\bm{\omega}T\Big(3\mu\partial_0\bar{\mu}+4I_1\partial_0T\Big)
+\frac{1}{2}I_1T^2\nabla\times(\partial_0{\bf u})
\Bigg],
\end{eqnarray}
where $I_{1,2\dots}$ are defined in Eqs.~(\ref{def_lambda})-(\ref{def_I22}). 
Next, we consider the contribution from $\delta f^{\mathcal{H}}_q$,
\begin{eqnarray}
\delta {\bf J}_{\mathcal{H}}&=&2\int\frac{d^4q}{(2\pi)^3}\bar{\epsilon}(q\cdot u)\delta(q^2){\bf q}\delta f^{\mathcal{H}}_q
=2\hbar\tau_R\int\frac{d^4q}{(2\pi)^3}\bar{\epsilon}(q_0)\delta(q^2)
\frac{q_0^2}{3}(\partial_0\delta {\bf u})\partial_{q0}f^{(0)}_q,
\end{eqnarray}
which results in  
\begin{eqnarray}
\delta {\bf J}_{\mathcal{H}}=-\frac{\hbar\tau_R}{6\pi^2}I_2T^3\Bigg(\tilde{U}_E\nabla\times {\bf E}+\tilde{U}_T\frac{{\bf E}\times\nabla T}{T}+\tilde{U}_{\bar{\mu}}{\bf E}\times\nabla\bar{\mu}\Bigg).
\end{eqnarray}
Subsequently, the third part stemming from collisions is given by
\begin{eqnarray}
\delta {\bf J}_{C}=2\int\frac{d^4q}{(2\pi)^3}\bar{\epsilon}(q\cdot u)\delta(q^2){\bf q}\delta f^{\mathcal{C}}_q
=\frac{2\hbar\tau_R\bm{\mathcal{A}}}{3}\int\frac{d^4q}{(2\pi)^3}\bar{\epsilon}(q_0)\delta(q^2)
\left(\frac{\partial_0T}{T}+\frac{T\partial_0\bar{\mu}}{q_0}\right)\partial_{q0}f^{(0)}_q.
\end{eqnarray} 
Performing the integral, we find
\begin{eqnarray}
\delta {\bf J}_{C}=-\frac{\hbar\tau_R\bm{\mathcal{A}}}{6\pi^2}\left(\frac{\mu\partial_0T}{T}+T\partial_0\bar{\mu}\right).
\end{eqnarray}
Note that $\delta {\bf J}_{C}$ vanishes after applying the hydrodynamic EOM in Eq.~(\ref{D0T_D0mu}).

Finally, we have to consider the contributions from $\delta f^{(c)}_q$ in Eq.~(\ref{delJQl}). We first evaluate the part related to the curl term and electric fields, which gives 
\begin{eqnarray}\nonumber
\delta {\bf J}_E
&=&-2\int\frac{d^4q}{(2\pi)^3}\bar{\epsilon}(q_0)\delta(q^2)\frac{\hbar}{2q_0}\Big({\bf q}\times\nabla+({\bf q\times E})\partial_{q0}-q_0{\bf E}\times\nabla_{\bf q}\Big)\delta f^{(c)}_q
\\\nonumber
&=&\frac{\hbar\tau_R}{3}\int\frac{d^4q}{(2\pi)^3}\bar{\epsilon}(q_0)\delta(q^2)
\Bigg[
\Bigg(-\nabla\times{\bf E}+\frac{2{\bf E}\times\nabla T}{T}+\frac{2T({\bf E}\times\nabla\bar{\mu})}{q_0}\Bigg)
+2q_0^2\frac{\partial_0T}{T}\bm{\omega}\partial_{q0}
\\
&&
+\Bigg(q_0\nabla\times\Big(\partial_0{\bf u}\Big)+{\bf E}\times\partial_0{\bf u}(3+q_0\partial_{q0})
+q_0(\partial_0{\bf u})\times\frac{\nabla T}{T}(1+q_0\partial_{q0})
\Bigg)
\Bigg]\partial_{q0}f^{(0)}_q.
\end{eqnarray}
Performing the integral, we obtain 
\begin{eqnarray}\nonumber
\delta {\bf J}_E&=&-\frac{\tau_R\hbar}{12\pi^2}\Bigg[
\Bigg(-\mu\nabla\times{\bf E}+2\bar{\mu}{\bf E}\times\nabla T+2T{\bf E}\times\nabla\bar{\mu}\Bigg)
-6I_1T\bm{\omega}(\partial_0T)
\\
&&
+\Bigg(\mu T\nabla\times\Big(\partial_0{\bf u}\Big)
+\mu {\bf E}\times\partial_0{\bf u}
-2I_1T(\partial_0{\bf u})\times\nabla T
\Bigg)
\Bigg].
\end{eqnarray}
Nonetheless, there exists a missing contribution in $\delta {\bf J}_E$ when considering just the ``naive'' RT approximation treating $\tau_R$ as a constant. In practice, $\tau_R$ should be a function of $T$ and $\mu$. The magnetization current stemming from ${\bf q}\times \nabla\delta f^{(c)}_q$ in the integrand could further gives rise to
\begin{eqnarray}\nonumber\label{delJ_tauR}
\delta {\bf J}_{\tau_R}&=&-2\int\frac{d^4q}{(2\pi)^3}\bar{\epsilon}(q_0)\delta(q^2)\frac{\hbar}{2q_0^2}{\bf q}\times(\nabla\tau_R)q\cdot\tilde{E}\partial_{q_0}f^{(0)}_q
\\\nonumber
&=&\frac{\hbar}{12\pi^2}\Big(I_1T\nabla T-\mu({\bf E}-T\nabla\bar{\mu})\Big)\times
\Big((\partial_T\tau_R)\nabla T+(\partial_{\bar{\mu}}\tau_R)\nabla\bar{\mu}\Big)
\\\nonumber
&=&\frac{\hbar}{12\pi^2}\Big((\mu\partial_T\tau_R-I_1\partial_{\bar{\mu}}\tau_R)(\nabla\mu)\times(\nabla T)+\bar{\mu}(\bar{\mu}\partial_{\bar{\mu}}\tau_R-T\partial_T\tau_R){\bf E}\times\nabla T
	-\bar{\mu}(\partial_{\bar{\mu}}\tau_R){\bf E}\times\nabla\mu
	\Big),
\\
\end{eqnarray}
which may generate ${\bf E}\times\nabla T$ and $(\nabla\mu)\times(\nabla T)$ terms even in the absence of hydrodynamics depending on the detailed structure of $\tau_R$. For making fair comparisons with previous studies, in which $\delta {\bf J}_{\tau_R}$ is ignored \cite{Chen:2016xtg,Gorbar:2016qfh}, we will apply the ``naive'' RT approximation and thus omit $\delta {\bf J}_{\tau_R}$. 
The rest part related magnetic fields reads
\begin{eqnarray}\nonumber
\delta {\bf J}_B&=&-2\int\frac{d^4q}{(2\pi)^3}\bar{\epsilon}(q_0)\delta(q^2)\frac{\hbar}{2q_0}\Bigg(q_0{\bf B}\partial_{q0}
-({\bf q\cdot B})\nabla_{\bf q}+{\bf B}({\bf q}\cdot\nabla_{\bf q})
\Bigg)
\delta f^{(0)}_q
\\\nonumber
&=&-\hbar\tau_R {\bf B}\int\frac{d^4q}{(2\pi)^3}\bar{\epsilon}(q_0)\delta(q^2)
\Bigg[\Big(T\partial_0\bar{\mu}+\frac{q_0\partial_0T}{T}\Big)\partial_{q0}
+\frac{\partial_0T}{T}
\Bigg]\partial_{q0}f^{(0)}_q,
\end{eqnarray}
which yields
\begin{eqnarray}
\delta {\bf J}_B
=-\frac{\tau_R\hbar}{4\pi^2}
{\bf B}\Big(T\partial_0\bar{\mu}+\bar{\mu}\partial_0T\Big).
\end{eqnarray}
Apparently, $\delta {\bf J}_B=0$ from hydrodynamics.

Combining all the contributions, we eventually obtain 
\begin{eqnarray}\nonumber
\delta {\bf J}_Q&=&\delta {\bf J}_{\mathcal{K}}+\delta {\bf J}_{\mathcal{H}}+\delta {\bf J}_{\mathcal{C}}+\delta{\bf J}_E+\delta {\bf J}_B
\\\nonumber
&=&-\frac{\hbar\tau_R}{12\pi^2}\Bigg[-2\Big(\mu-I_2T^3\tilde{U}_{E}\Big)\nabla\times{\bf E}
+\Big(1+2I_2T^2\tilde{U}_{\bar{\mu}}\Big) {\bf E}\times\nabla\mu
\\\nonumber
&&+2I_2T\Big(T\tilde{U}_T-\mu\tilde{U}_{\bar{\mu}}\Big){\bf E}\times\nabla T
+T\Big(2\bar{\mu}{\bf E}-\mu\nabla\bar{\mu}+I_1\nabla T\Big)\times\partial_0{\bf u}
+\frac{3I_1T^2}{2}\nabla\times(\partial_0{\bf u})
\\
&&+3{\bf B}\Big(T\partial_0\bar{\mu}+\bar{\mu}\partial_0T\Big)
-\bm{\omega}T\Big(2I_1\partial_0T-3\mu\partial_0\bar{\mu}\Big)
+2\bm{\mathcal{A}}\Big(\bar{\mu}\partial_0T+T\partial_0\bar{\mu}\Big)
\Bigg].
\end{eqnarray}
By omitting the terms including time derivatives and taking $\tilde{U}_{E,T,\bar{\mu}}=0$,
it is clear to see that only the ${\bf\nabla\times E}$ and ${\bf E}\times\nabla\mu$ terms contribute to the non-equilibrium charge current in the absence of hydrodynamics, which agrees with what have been found in Refs.~\cite{Chen:2016xtg,Gorbar:2016qfh}. Now, by using hydrodynamic EOM in Eqs.~(\ref{D0T_D0mu}) and (\ref{D0ui_full}) to replace the time derivatives of $T$, $\bar{\mu}$, and ${\bf u}$ with explicit expressions, the non-equilibrium charge current becomes 
\begin{eqnarray}\nonumber\label{charge_J_hydro}
\delta {\bf J}_Q&=&-\frac{\hbar\tau_R}{12\pi^2}\Bigg[-2\left(\mu-I_2T^3\tilde{U}_{E}-\frac{3I_1T^2N_0}{16p}\right)\nabla\times{\bf E}+\left(\bar{\mu}+\frac{I_1TN_0}{4p}\right)(\nabla\mu)\times(\nabla T)
\\\nonumber
&&+\left(\left(1-\frac{\mu N_0}{4p}\right)
+2I_2T^2\tilde{U}_{\bar{\mu}}+\frac{3I_1T^2}{4p}\left(\frac{N_0^2}{2p}+\sigma_{\omega}\right)\right){\bf E}\times\nabla\mu
\\
&&
+\Bigg(2I_2T(T\tilde{U}_T-\mu\tilde{U}_{\bar{\mu}})-2\bar{\mu}
+\frac{\mu^2 N_0}{4Tp}
+\frac{I_1T}{8p}\left(N_0-6\mu\sigma_{\omega}-\frac{3\mu N_0^2}{p}\right)
\Bigg){\bf E}\times\nabla T
\Bigg],
\end{eqnarray}
where we utilize
\begin{eqnarray}
\frac{1}{2}\nabla\times(\partial_0{\bf u})
=\frac{1}{8p}
\Bigg(\frac{N_0{\bf E}\times\nabla T}{T}+N_0\nabla\times{\bf E}
+T\left(\frac{N_0^2}{p}-2\sigma_{\omega}\right){\bf E}\times\nabla\bar{\mu}\Bigg)+\mathcal{O}(\hbar)
\end{eqnarray}
from hydrodynamic EOM in the calculation. Except for the modifications of the transport coefficient for ${\bf\nabla\times E}$ and ${\bf E}\times\nabla\mu$ terms, ${\bf E}\times\nabla T$ and $(\nabla \mu)\times(\nabla T)$ terms emerge in hydrodynamics. Note that here the vorticity does not affect $\delta {\bf J}_Q$ in the inviscid case. Also, in contrast to the anomalous Hall current above from quantum corrections, the classical Hall current obtained from the kinetic theory in weak fields and the RT approximation is proportional to $\tau_R^2$ instead of $\tau_R$ as shown in e.g., Refs.~\cite{Pu:2014fva, Gorbar:2016qfh}.

\subsection{Charge Density} \label{charge density}
Following Eq.~(\ref{delJQl}), we can further evaluate the charge density. Similar to the computation for currents, we first consider the contribution led by $\delta f^{\mathcal{K}}_q$, which yields
\begin{eqnarray}\nonumber
\delta J_{\mathcal{K}}^0&=&2\int\frac{d^4q}{(2\pi)^3}\bar{\epsilon}(q\cdot u)\delta(q^2)q^{0}\delta f^{\mathcal{K}}_q
\\\nonumber
&=&2\tau_R\int\frac{d^4q}{(2\pi)^3}\bar{\epsilon}(q_0)\delta(q^2)
\frac{\hbar}{2}\Bigg[-\bm{\omega}\cdot\bm{\mathcal{E}}+\frac{q_0\bm{\omega}\cdot\nabla T}{T}
+\frac{(\omega\cdot \mathcal{E})}{3}
+\frac{q_0(\nabla\cdot\bm{\omega})}{3}-q_0\partial_0\omega^0
\\
&&+\frac{q_0}{3}\Big(\bm{\omega}\cdot\bm{\mathcal{E}}-\frac{q_0\bm{\omega}\cdot\nabla T}{T}\Big)\partial_{q0}
+\frac{q_0}{3}\bm{\omega}\cdot\partial_0{\bf u}(2-\partial_{q0})+q_0\bm{\omega}\cdot\partial_0{\bf u}
\Bigg]\partial_{q0}f^{(0)}_q.
\end{eqnarray}
By employing the Bianchi identity for vorticity in Eq.~(\ref{domega2}) and performing the integral, we obtain
\begin{eqnarray}
\delta J_{\mathcal{K}}^0
=\frac{\hbar\tau_R}{2\pi^2}\Bigg[ \mu\bm{\omega}\cdot\bm{\mathcal{E}}-I_1T\bm{\omega}\cdot\nabla T
-I_1T^2\bm{\omega}\cdot\partial_0{\bf u}
\Bigg].
\end{eqnarray}
Next, considering the quantum corrections from hydrodynamic EOM, we find
\begin{eqnarray}\nonumber
\delta J_{\mathcal{H}}^0&=&2\int\frac{d^4q}{(2\pi)^3}\bar{\epsilon}(q\cdot u)\delta(q^2)q^{0}\delta f^{\mathcal{H}}_q
\\
&=&2\hbar\tau_R\int\frac{d^4q}{(2\pi)^3}\bar{\epsilon}(q\cdot u)\delta(q^2)
q_0\bm{\mathcal{E}}\cdot\Bigg[ T\Big(\tilde{\mu}_B{\bf B}+\tilde{\mu}_{\omega}\bm{\omega}T\Big)+q_0\Big(\tilde{T}_B{\bf B}+\tilde{T}_{\omega}\bm{\omega}T\Big)\Bigg]\partial_{q0}f^{(0)}_q,
\end{eqnarray}
which results in
\begin{eqnarray}
\delta J_{\mathcal{H}}^0
=-\frac{\hbar\tau_R}{2\pi^2}\Big[\Big(\bm{\mathcal{E}}\cdot {\bf B}\Big)T^3\Big(I_1\tilde{\mu}_B+I_2\tilde{T}_B\Big)
+\Big(\bm{\mathcal{E}}\cdot \bm{\omega}\Big)T^4\Big(I_1\tilde{\mu}_{\omega}+I_2\tilde{T}_{\omega}\Big)
\Big].
\end{eqnarray}
As opposed to the case for the charge current, the relevant parts in hydrodynamic EOM are $\partial_0T$ and $\partial_0\bar{\mu}$. For the charge density, there is only magnetic-field related term coming from the side-jump term associated with magnetization currents as shown in Eq.~(\ref{delJQl}). We thus evaluate
\begin{eqnarray}\nonumber
\delta J^0_{\mathcal{B}}=&=&2\int\frac{d^4q}{(2\pi)^3}\bar{\epsilon}(q_0)\delta(q^2)\frac{\hbar}{2}
{\bf B}\cdot\nabla_{\bf q}
\delta f^{(0)}_q
\\
&=&2\tau_R\int\frac{d^4q}{(2\pi)^3}\bar{\epsilon}(q_0)\delta(q^2)\frac{\hbar}{2q_0}
\Big(-{\bf B}\cdot \bm{\mathcal{E}}+q_0\frac{{\bf B}\cdot \nabla T}{T}+q_0{\bf B}\cdot\partial_0{\bf u}\Big)\partial_{q0}f^{(0)}_q
\end{eqnarray}
and derive
\begin{eqnarray}
\delta J^0_{\mathcal{B}}
=-\frac{\hbar\tau_R}{2\pi^2}
\Big(-({\bf B}\cdot \bm{\mathcal{E}})+\bar{\mu}{\bf B}\cdot\nabla T+\mu {\bf B}\cdot\partial_0{\bf u}\Big).
\end{eqnarray}
Finally, considering the $\hbar$ corrections in collisions, we find
\begin{eqnarray}\nonumber
\delta J_{C}^0&=&2\int\frac{d^4q}{(2\pi)^3}\bar{\epsilon}(q\cdot u)\delta(q^2)q^{0}\delta f^{\mathcal{C}}_q
\\
&=&2\hbar\tau_R\int\frac{d^4q}{(2\pi)^3}\bar{\epsilon}(q_0)\delta(q^2)
\frac{1}{3q_0}\Bigg[-\bm{\mathcal{A}}\cdot\bm{\mathcal{E}}+q_0\frac{\bm{\mathcal{A}}\cdot\nabla T}{T}+q_0\bm{\mathcal{A}}\cdot\partial_0{\bf u}\Bigg]\partial_{q0}f^{(0)}_q,
\end{eqnarray}
which gives
\begin{eqnarray}\label{J0Q}
\delta J_{C}^0=-\frac{\hbar\tau_R}{6\pi^2}\Bigg[-\bm{\mathcal{A}}\cdot\bm{\mathcal{E}}+\bar{\mu}\bm{\mathcal{A}}\cdot\nabla T+\mu\bm{\mathcal{A}}\cdot\partial_0{\bf u}\Bigg].
\end{eqnarray}

Combining all pieces together and utilizing the hydrodynamic EOM, it turns out that except for $\delta J^0_C$, the nonlinear quantum corrections on charge density vanishes,
\begin{eqnarray}\label{zero_c_density}
\delta J_{\mathcal{K}}^0+\delta J_{\mathcal{H}}^0+\delta J_{\mathcal{B}}^0=0.
\end{eqnarray}
Nonetheless, the finding is not surprising, which in fact agrees with the matching condition of the RT approximation as shown below. As found in Appendix~\ref{sec:2ndOrder}, the CKT gives rise to 
\begin{eqnarray}\label{anomalous_eq}
  \partial_{\mu}J^{\mu}=\frac{\hbar}{4\pi^2}({\bf E\cdot B})+2\int\frac{d^4q}{(2\pi)^3}\bar{\epsilon}(q\cdot u)\left[\delta(q^2)q\cdot\tilde{\mathcal{C}}+\hbar\epsilon^{\mu\nu\alpha\beta}\mathcal{C}_{\mu}F_{\alpha\beta}\frac{\partial_{q\nu}\delta(q^2)}{4}\right].
\end{eqnarray}
For realistic collisions, the integral with collisional terms should automatically vanish in accordance with energy-momentum conservation. In the case for $\tilde{\mathcal{C}}_i=\mathcal{C}_{i}=0$, we may rewrite the last term as
\begin{eqnarray}\nonumber\label{matching_cond}
&&2\int\frac{d^4q}{(2\pi)^3}\bar{\epsilon}(q\cdot u)\left[\delta(q^2)q\cdot\tilde{\mathcal{C}}+\hbar\epsilon^{\mu\nu\alpha\beta}\mathcal{C}_{\mu}F_{\alpha\beta}\frac{\partial_{q\nu}\delta(q^2)}{4}\right]
\\
&&=2\int\frac{d^4q}{(2\pi)^3}\bar{\epsilon}(q_0)\delta(q^2)q_0\Bigg(1+\frac{\hbar {\bf B}\cdot\nabla_{\bf q}}{2q_0}\Bigg)\mathcal{C}_0.
\end{eqnarray}
Applying the RT approximation in Eq.~(\ref{RT_approx}) for $q_0\mathcal{C}_0=\mathcal{C}_\text{full}$, such a vanishing term results in the matching condition
\begin{eqnarray}\nonumber
&&2\int\frac{d^4q}{(2\pi)^3}\bar{\epsilon}(q\cdot u)\delta(q^2)\left(1+\frac{\hbar {\bf B}\cdot\nabla_{\bf q}}{2q_0}\right)\mathcal{C}_\text{full}
\\\nonumber
&&=-2\int\frac{d^4q}{(2\pi)^3}\bar{\epsilon}(q_0)\delta(q^2)\frac{q_0}{\tau_R}\left(\delta f_q^{(c)}+\frac{\hbar {\bf B}\cdot\nabla_{\bf q}}{2q_0}\delta f_q^{(c)}+\delta f^{\mathcal{K}}_q+\delta f^{\mathcal{H}}_q\right)
\\
&&=0.
\end{eqnarray}
Note that the ${\bf q}\cdot\bm{\mathcal{A}}$ term in the RT approximation and that in $\delta f^{C}_q$ cancel each other.
Apparently, this matching condition in general leads to the vanishing charge density from interactions $J^0_\text{int}=0$. One can easily check that the classical contribution above is zero. Then the vanishing quantum correction is consistent with (\ref{zero_c_density}). Nevertheless, the matching condition here does not exclude the contribution from $\delta f^{C}_q$ for the charge density. 
 
\subsection{Vector/Axial-Charge Currents}\label{va_currents}
So far, we have considered only the responses for right-handed fermions. One can implement the same approach to derive the responses for left-handed fermions, for which the $\hbar$ corrections are the same as those for right-handed fermions with just an overall sign difference. Combining the contributions from both right-handed and left-handed fermions, we can compute the vector/axial charge currents. Due to the complexity of transport coefficients in hydrodynamics, we may focus on just high-temperature and large-chemical-potential limits. In addition, only the quantum corrections will be presented. One may refer to Ref.~\cite{Gorbar:2016qfh} for a complete study with the classical second-order effects in vector/axial charge currents in the absence of collective motion. 

From Eq.~(\ref{charge_J_hydro}), in the high-temperature limit $\bar{\mu}_{R/L}\ll 1$, the right/left-handed charge currents are given by 
\begin{eqnarray}\label{charge_J_R_highT}\nonumber
\delta {\bf J}_{QR/L}=\mp\frac{\hbar\tau_R}{84\pi^2}\Bigg[\mu_{R/L}\nabla\times{\bf E}+\frac{7\mu_{R/L}}{2T}{\bf E}\times\nabla T-\frac{1}{2}{\bf E}\times\nabla \mu_{R/L}+\frac{12\mu_{R/L}}{T}(\nabla\mu_{R/L})\times(\nabla T)
\Bigg],
\\
\end{eqnarray}
where we append the subindices $R/L$ to $\mu$ for representing right/left-handed chemical potentials and the overall minus/plus signs are for right/left-handed fermions, respectively. Here we preserve the leading-order contribution in terms of the small-$\bar{\mu}_{R/L}$ expansion for each relevant term. On the contrary, for $\bar{\mu}_{R/L}\gg 1$, the right/left-handed charge currents reduce to  
\begin{eqnarray}\label{charge_J_R_lowT}\nonumber
\delta {\bf J}_{QR/L}=\pm\frac{\hbar\tau_R}{4\pi^2}\Bigg[\frac{\pi^2 T^2}{3\mu_{R/L}}\nabla\times{\bf E}+\frac{2\mu_{R/L}}{3T}{\bf E}\times\nabla T-\frac{2}{3}{\bf E}\times\nabla \mu_{R/L}-\frac{2\mu_{R/L}}{3T}(\nabla\mu_{R/L})\times(\nabla T)
\Bigg],
\\
\end{eqnarray}  
via the $1/\bar{\mu}_{R/L}$ expansion.
Here some of transport coefficients change signs in different limits, for which the reason in underlying physics is unclear. However, as mentioned previously, in more pragmatic cases, one should also incorporate the contributions from $\delta {\bf J}_{\tau_R}$ shown in Eq.~(\ref{delJ_tauR}). In general, to pin down the signs and numerical values of these transport coefficients, we certainly have to work beyond the RT approximation, whereas their dependence on $T$ and $\mu_{R/L}$ could be  qualitatively captured by the results above.   

We may now introduce the vector/axial-charge currents, ${\bf J}_{V/A}={\bf J}_R\pm {\bf J}_L$. Considering both the right/left-handed fermions, we obtain, for high temperature $\bar{\mu}_{R/L}\ll 1$, 
\begin{eqnarray}\nonumber\label{charge_J_VA_highT}
\delta {\bf J}_{QV}&=&-\frac{\hbar\tau_R}{84\pi^2}\Bigg[\mu_A\nabla\times{\bf E}+\frac{7\mu_A}{2T}{\bf E}\times\nabla T-\frac{1}{2}{\bf E}\times\nabla\mu_A+\frac{6}{T}\nabla(\mu_V\mu_A)\times(\nabla T)
\Bigg],
\\
\delta {\bf J}_{QA}&=&-\frac{\hbar\tau_R}{84\pi^2}\Bigg[\mu_V\nabla\times{\bf E}+\frac{7\mu_V}{2T}{\bf E}\times\nabla T-\frac{1}{2}{\bf E}\times\nabla\mu_V+\frac{3}{T}\nabla(\mu_V^2+\mu_A^2)\times(\nabla T)
\Bigg],
\end{eqnarray}
where $\mu_{V/A}=\mu_R\pm\mu_L$. Note that in the extreme case with zero net chemical potentials, only the ${\bf E}\times \nabla\mu_{A/V}$ terms contribute to the vector/axial currents. In the large-chemical-potential limit $\bar{\mu}_{R/L}\gg 1$, one finds 
\begin{eqnarray}\nonumber\label{charge_J_VA_lowT}
\delta {\bf J}_{QV}&=&\frac{\hbar\tau_R}{4\pi^2}\Bigg[\frac{4\pi^2 T^2\mu_A}{3(\mu_A^2-\mu_V^2)}\nabla\times{\bf E}+\frac{2\mu_A}{3T}{\bf E}\times\nabla T-\frac{2}{3}{\bf E}\times\nabla\mu_A-\frac{1}{3T}\nabla(\mu_V\mu_A)\times(\nabla T)
\Bigg],
\\
\delta {\bf J}_{QA}&=&\frac{\hbar\tau_R}{4\pi^2}\Bigg[\frac{4\pi^2 T^2\mu_V}{3(\mu_V^2-\mu_A^2)}\nabla\times{\bf E}+\frac{2\mu_V}{3T}{\bf E}\times\nabla T-\frac{2}{3}{\bf E}\times\nabla\mu_V-\frac{1}{6T}\nabla(\mu_V^2+\mu_A^2)\times(\nabla T)
\Bigg].
\end{eqnarray} 
As opposed to the high-temperature limit, here the dominant contributions in vector/axial charge currents come from ${\bf E}\times\nabla T$ and $(\nabla\mu_{V/A})\times(\nabla T)$ terms.

\section{Discussions and Outlook}\label{Discussions_conclusions}
In this paper, we have investigated the dissipative quantum transport of inviscid chiral fluids incorporating background fields and vorticity up to the second order via the CKT with a RT approximation. It is found that some of anomalous Hall currents emerge from the anomalous-hydrodynamic corrections. Furthermore, we show that the vanishing quantum correction upon the charge density (except for the potential contributions from quantum corrections in collisions) agrees with the matching condition obtained from CKT based on the energy-momentum conservation. 

Although we find extra Hall currents coming from the cross product of electric fields and temperature-gradient and that of the temperature and chemical-potential gradients given by anomalous hydrodynamic corrections, we also indicate that such terms could possibly led by the temperature/chemical-potential dependence of relaxation time even in the absence of hydrodynamics. In phenomenology, this implies that the aforementioned anomalous Hall currents could possibly exist in Weyl semimetals even in the absence of collective motion for quasi-particles. The study along such a direction will be presented elsewhere. On the other hand, the possible quantum corrections in collisions stemming from side jumps may contribute to the charge density. In fact, for realistic collisions, the anomalous equation does not force the correction on the charge density to be always zero. It is found in a recent study that the direct product of magnetic fields and vorticity can modify the charge density for Weyl fermions in the lowest Landau level \cite{Hattori:2016njk}. However, we suspect such an effect may be at $\mathcal{O}(\hbar^2)$ in the Wigner-function approach with weak background fields. In addition, even though the quantum corrections in collisions characterized by the RT approximation does not influence the charge current in inviscid fluids, this may not be the case for viscous ones. To further explore the nonlinear quantum transport for viscous chiral fluids, the current RT approximation may be insufficient. Either directly tackling realistic yet complicated collisional kernels including side-jump effects or developing more applicable approximations for collisions should be pursued in the future.           
 
\appendix

\section{Conventions and Useful Relations \label{convention}} 
In this paper, we use the most negative spacetime metric $\eta_{\mu\nu}=\text{diag}(1,-1,-1,-1)$ and we define $\epsilon^{0ijk}=-\epsilon_{0ijk}\equiv\epsilon^{ijk}$ with $\epsilon^{123}=1$.
Also, $\bar{\mu}=\mu/T$ and $n^{\mu}$ denotes the frame vector and $u^{\mu}$ represents the fluid velocity.

The fluid vorticity is defined as 
\begin{eqnarray}
\omega^{\mu}\equiv\frac{T}{2}\epsilon^{\mu\nu\alpha\beta}u_{\nu}\partial_{\alpha}(\beta u_{\beta})=\frac{1}{2}\epsilon^{\mu\nu\alpha\beta}u_{\nu}\big(\partial_{\alpha}u_{\beta}\big).
\end{eqnarray}
We apply the decompositions 
\begin{eqnarray}
\partial_{\mu}u_{\nu}=\sigma_{\mu\nu}+\omega_{\mu\nu},\quad\sigma_{\mu\nu}=\frac{1}{2}\big(\partial_{\mu}u_{\nu}+\partial_{\nu}u_{\mu}\big),\quad\omega_{\mu\nu}=\frac{1}{2}\big(\partial_{\mu}u_{\nu}-\partial_{\nu}u_{\mu}\big)
\end{eqnarray}
and 
\begin{eqnarray}
\omega_{\alpha\beta}=-\epsilon_{\alpha\beta\mu\nu}\omega^{\mu}u^{\nu}+\kappa_{\alpha\beta},\quad\kappa_{\alpha\beta}=\frac{1}{2}\big(u_{\alpha}u\cdot\partial u_{\beta}-u_{\beta}u\cdot\partial u_{\alpha}\big),
\end{eqnarray}
which gives 
\begin{eqnarray}
T\partial_{\mu}(\beta u_{\nu})=-\epsilon_{\mu\nu\alpha\beta}u^{\beta}\omega^{\alpha}+\kappa_{\mu\nu}+\sigma_{\mu\nu}-\frac{u_{\nu}\partial_{\mu}T}{T}.
\end{eqnarray}
We also introduce the dual tensor 
\begin{eqnarray}
\tilde{\omega}^{\mu\nu}=\frac{1}{2}\epsilon^{\mu\nu\alpha\beta}\omega_{\alpha\beta}=\frac{1}{2}\epsilon^{\mu\nu\alpha\beta}u_{\alpha}u\cdot\partial u_{\beta}+\omega^{\mu}u^{\nu}-\omega^{\nu}u^{\mu},\quad\omega_{\mu\nu}=-\frac{1}{2}\epsilon_{\mu\nu\alpha\beta}\tilde{\omega}^{\alpha\beta}.
\end{eqnarray}
In the $n^{\mu}$ frame, the electromagnetic fields are given by 
\begin{eqnarray}
n^{\nu}F_{\mu\nu}=E_{\mu},\quad\frac{1}{2}\epsilon^{\mu\nu\alpha\beta}n_{\nu}F_{\alpha\beta}=B^{\mu},\quad F_{\alpha\beta}=-\epsilon_{\mu\nu\alpha\beta}B^{\mu}n^{\nu}+n_{\beta}E_{\alpha}-n_{\alpha}E_{\beta}.
\end{eqnarray}
When $n^{\mu}=u^{\mu}$, the Bianchi identity $\partial_{\nu}\tilde{F}^{\mu\nu}=0$
yields
\begin{eqnarray}
\partial\cdot B-2E\cdot\omega+B^{\mu}u\cdot\partial u_{\mu}=0,
\end{eqnarray}
and 
\begin{eqnarray}
u\cdot\partial B^{\rho}+B^{\rho}\partial\cdot u-B\cdot\partial u^{\rho}+u^{\rho}B^{\mu}u\cdot\partial u_{\mu}+\epsilon^{\rho\mu\alpha\beta}\big(u_{\beta}\partial_{\mu}E_{\alpha}+u_{\mu}E_{\alpha}u\cdot\partial u_{\beta}\big)=0.\label{eq:dB1}
\end{eqnarray}
In the local rest frame, it becomes 
\begin{eqnarray}
 &  & \partial\cdot B+2{\bf E}\cdot\bm{\omega}-{\bf B}\cdot\partial_{0}{\bf u}=0,\nonumber \\
 &  & \partial_{0}{\bf B}+{\bf B}(\partial\cdot u)-\frac{1}{2}\big({\bf (B\cdot\nabla)u}+{\bf B}\cdot(\nabla{\bf u})\big)+{\bf B\times\bm{\omega}}+(\nabla\times{\bf E}-{\bf E}\times\partial_{0}{\bf u})=0.
\end{eqnarray}
In the paper, we also widely use the notation 
\begin{eqnarray}
\tilde{E}_{\beta}=\mathcal{E}_{\beta}+\frac{(q\cdot u)}{T}\partial_{\beta}T-q^{\sigma}(\sigma_{\beta\sigma}+\kappa_{\beta\sigma}),\quad\mathcal{E}_{\mu}=E_{\mu}+T\partial_{\mu}\bar{\mu}.
\end{eqnarray}

Similar to $F_{\mu\nu}$, $\partial_{\nu}\tilde{\omega}^{\mu\nu}=0$
gives (using $\omega^{\mu}\rightarrow-B^{\mu}$, $\frac{u\cdot\partial}{2}u_{\beta}\rightarrow-E_{\beta}$)
\begin{eqnarray}
\partial\cdot\omega+2\omega^{\mu}u\cdot\partial u_{\mu}=0,
\end{eqnarray}
and 
\begin{eqnarray}
u\cdot\partial\omega^{\rho}+\omega^{\rho}\partial\cdot u-\omega\cdot\partial u^{\rho}+u^{\rho}\omega^{\mu}u\cdot\partial u_{\mu}+\frac{\epsilon^{\rho\mu\alpha\beta}}{2}u_{\beta}\partial_{\mu}(u\cdot\partial u_{\alpha})=0.\label{eq:domega1}
\end{eqnarray}
In the local rest frame, we accordingly have 
\begin{eqnarray}\label{domega2}
\nabla\cdot\bm{\omega}
=0,\quad\partial_{0}\bm{\omega}+\bm{\omega}\partial\cdot u=\frac{1}{2}\Big[{\bf (\bm{\omega}\cdot\nabla)u}+\bm{\omega}\cdot(\nabla{\bf  u})-\nabla\times(\partial_{0}{\bf u})\Big].
\end{eqnarray}

There exist useful integrals : 
\begin{eqnarray}
-\int\frac{d^{4}q}{(2\pi)^{3}}\theta(q_{0})\delta(q^{2})q_{0}^{-1}\partial_{q0}f^{(0)}=\frac{1}{4\pi^{2}},\label{def_lambda}
\end{eqnarray}
\begin{eqnarray}
I_{0}=-(4\pi^2)T^{-1}\int\frac{d^{4}q}{(2\pi)^{3}}\theta(q_{0})\delta(q^{2})\partial_{q0}f^{(0)}=\bar{\mu},
\end{eqnarray}
\begin{eqnarray}
I_{1}=-(4\pi^2)T^{-2}\int\frac{d^{4}q}{(2\pi)^{3}}\theta(q_{0})\delta(q^{2})q_{0}\partial_{q0}f^{(0)}=\bar{\mu}^{2}+\frac{\pi^{2}}{3},
\end{eqnarray}
\begin{eqnarray}
I_{2}=-(4\pi^2)T^{-3}\int\frac{d^{4}q}{(2\pi)^{3}}\theta(q_{0})\delta(q^{2})q_{0}^{2}\partial_{q0}f^{(0)}=\bar{\mu}\big(\bar{\mu}^{2}+\pi^{2}\big),
\end{eqnarray}
\begin{eqnarray}
I_{3}=-(4\pi^2)T^{-4}\int\frac{d^{4}q}{(2\pi)^{3}}\theta(q_{0})\delta(q^{2})q_{0}^{3}\partial_{q0}f^{(0)}=\frac{7\pi^{4}}{15}+2\pi^{2}\bar{\mu}^{2}+\bar{\mu}^{4},
\end{eqnarray}
\begin{eqnarray}
I_{1}^{(2)}=-(4\pi^2)T^{-1}\int\frac{d^{4}q}{(2\pi)^{3}}\theta(q_{0})\delta(q^{2})q_{0}\partial_{q0}^{2}f^{(0)}=-2I_{0},
\end{eqnarray}
\begin{eqnarray}\label{def_I22}
I_{2}^{(2)}=-(4\pi^2)T^{-2}\int\frac{d^{4}q}{(2\pi)^{3}}\theta(q_{0})\delta(q^{2})q_{0}^{2}\partial_{q0}^{2}f^{(0)}=-3I_{1}
\end{eqnarray}

\section{Covariant Anomalous Hydrodynamic Equations \label{Ahydro}}

From Eq.~(\ref{delta_TEM}), we obtain,
\begin{equation}
\Theta^{\nu\mu}\partial_{\mu}\frac{1}{T}+\frac{1}{T}(u\cdot\partial)u^{\nu}+\frac{\hbar}{T(\epsilon+p)} \Xi^{\nu}-\frac{\Theta^{\nu\mu}}{\epsilon+p}\left[N_{0}\left(\partial_{\mu}\frac{\mu}{T}+\frac{E_{\mu}}{T}\right)+\xi\frac{1}{T}\epsilon^{\mu\nu\alpha\beta}\omega_{\nu}u_{\alpha}B_{\beta}\right]=0,\label{eq:temp_identity_ideal_fluid}
\end{equation}
with 
\begin{eqnarray}
\Xi^{\mu} & = & \xi_{B}B^{\mu}(\partial\cdot u)+\Theta^{\mu\alpha}(u\cdot\partial)(\xi_{B}B_{\alpha})+\xi_{B}(B\cdot\partial)u^{\mu}\nonumber \\
 &  & +\xi_{\omega}\omega^{\mu}(\partial\cdot u)+\Theta^{\mu\alpha}(u\cdot\partial)(\xi_{\omega}\omega_{\alpha})+\xi_{\omega}(\omega\cdot\partial)u^{\mu}.
\end{eqnarray}
From Eq.~(\ref{coefficients1}), the equation of states in this case
is,

\begin{equation}
\epsilon=3p.
\end{equation}
We choose $\beta=1/T$ and $\bar{\mu}$ as thermal variables and rewrite
the thermodynamical relations, 
\begin{eqnarray}
\frac{1}{3}\beta d\epsilon & = & \beta dp=-(\epsilon+p)d\beta+N_{0}d\bar{\mu},\nonumber \\
d\left(\beta p\right) & = & -\epsilon d\beta+N_{0}d\bar{\mu},\label{eq:thermal-temp-01}
\end{eqnarray}
where  
\begin{equation}
\epsilon=\left.\frac{\partial(\beta p)}{\partial\beta}\right|_{\bar{\mu}},\;N_{0}=\left.\frac{\partial(\beta p)}{\partial\bar{\mu}}\right|_{\beta}.
\end{equation}
Since $d(\beta p)$ is a total derivative, then we obtain, 
\begin{eqnarray}
\left.\frac{\partial N_{0}}{\partial\beta}\right|_{\bar{\mu}} & = & -\left.\frac{\partial\epsilon}{\partial\bar{\mu}}\right|_{\beta,}=-3N_{0}T,\nonumber \\
\label{eq:thermal-temp_02}
\end{eqnarray}

For convenience, we also derivative some useful relations, 
\begin{eqnarray}
\partial_{T}\sigma_{\omega}=\frac{2\sigma_{\omega}}{T}, & \quad & \partial_{\bar{\mu}}\sigma_{\omega}=2\xi_{B}T,\nonumber \\
\partial_{T}\sigma_{B}=\frac{\sigma_{B}}{T}, & \quad & \partial_{\bar{\mu}}\sigma_{B}=-CT,\nonumber \\
\partial_{T}\xi_{\omega}=\frac{3N_{0}}{T}, & \quad & \partial_{\bar{\mu}}\xi_{\omega}=2\sigma_{\omega}T,\nonumber \\
\partial_{T}\xi_{B}=\frac{\sigma_{\omega}}{T}, & \quad & \partial_{T}\xi_{B}=\sigma_{B}T,\nonumber \\
\partial_{T}\epsilon=\frac{4\epsilon}{T}, & \quad & \partial_{\bar{\mu}}\epsilon=3N_{0}T,
\end{eqnarray}
with $C=-\frac{1}{4\pi^{2}}$. Then, Eq.~(\ref{eq:temp_identity_ideal_fluid})
reduces to,
\begin{eqnarray}
(u\cdot\partial)u^{\nu} & = & \frac{N_{0}T}{\epsilon+p}
\mathcal{E}^{\nu}+\frac{1}{T}\nabla_{\mu}T+\hbar\frac{1}{\epsilon+p}{\frac{1}{2}}\sigma_{\omega}\epsilon^{\nu\rho\alpha\beta}\omega_{\rho}u_{\alpha}B_{\beta}\nonumber \\
 &  & -\frac{\hbar}{(\epsilon+p)}\left[\left(\frac{\sigma_{\omega}}{T}(u\cdot\partial)T
+\sigma_{B}T(u\cdot\partial)\bar{\mu}\right)B^{\nu}+\frac{1}{2}\sigma_{\omega}(u\cdot\partial)B^{\nu}+\frac{1}{2}\sigma_{\omega}(B^{\alpha}(u\cdot\partial)u_{\alpha})u^{\nu}\right]\nonumber \\\nonumber
 &  & -\frac{\hbar}{(\epsilon+p)}\left[\left(\frac{3N_{0}}{T}(u\cdot\partial)T+2\sigma_{\omega}T(u\cdot\partial)\bar{\mu}\right)\omega^{\nu}+N_{0}(u\cdot\partial)\omega^{\nu}+N_{0}(\omega^{\alpha}(u\cdot\partial)u_{\alpha})u^{\nu}\right],
 \\\label{eq:Du_1}
\end{eqnarray}
where $\nabla_{\mu}=\Theta_{\mu\nu}\partial^{\nu}$. Inserting $(u\cdot\partial)u^{\nu}$
into Eqs. (\ref{eq:dB1}) and (\ref{eq:domega1}), we get
\begin{eqnarray}
(u\cdot\partial)B^{\nu} & = & \epsilon^{\mu\nu\alpha\beta}(\partial_{\mu}E_{\alpha})u_{\beta}+(-u^{\nu}B^{\sigma}+\epsilon^{\nu\alpha\rho\sigma}E_{\alpha}u_{\rho})\left(\frac{N_{0}T}{\epsilon+p}\mathcal{E}_{\sigma}+\frac{1}{T}\nabla_{\sigma}T\right)\nonumber \\
 &  & +\epsilon^{\mu\nu\alpha\beta}B_{\mu}u_{\alpha}\omega_{\beta},\label{eq:DB_1}
\end{eqnarray}
and, 
\begin{eqnarray}
(u\cdot\partial)\omega^{\mu} & = & -\omega^{\nu}\left(\frac{N_{0}T}{\epsilon+p}\mathcal{E}_{\nu}+\frac{1}{T}\nabla_{\nu}T\right)u^{\mu}+\frac{3}{8}\frac{T}{\epsilon}\left(2\sigma_{\omega}-\frac{3N_{0}^{2}}{\epsilon}\right)\epsilon^{\mu\nu\alpha\beta}u_{\nu}(\partial_{\alpha}\bar{\mu})E_{\beta}-\frac{N_{0}T}{\epsilon+p}\omega^{\mu}(u\cdot \partial)\bar{\mu}\nonumber \\
 &  & -\frac{1}{2}\frac{N_{0}}{T(\epsilon+p)}\epsilon^{\mu\nu\alpha\beta}u_{\nu}(\partial_{\alpha}T)E_{\beta}+\frac{1}{2}\frac{N_{0}}{\epsilon+p}\epsilon^{\mu\nu\alpha\beta}u_{\nu}\partial_{\alpha}E_{\beta}-\frac{1}{T}\omega^{\mu}(u\cdot \partial)T,\label{eq:Domega_1}
\end{eqnarray}
where we have used, 
\begin{eqnarray}
(u\cdot\partial)(\partial_{\alpha}u_{\beta}) & = & \partial_{\alpha}((u\cdot\partial)u_{\beta})-\Theta_{\alpha\beta}\omega^{2}+\omega_{\alpha}\omega_{\beta}{\color{blue}+}u_{\alpha}\epsilon_{\beta\gamma\rho\sigma}((u\cdot\partial)u^{\gamma})u^{\rho}\omega^{\sigma},
\end{eqnarray}
and,
\begin{equation}
\partial_{T}\left(\frac{N_{0}T}{\epsilon+p}\right)=0,\;\partial_{\bar{\mu}}\left(\frac{N_{0}T}{\epsilon+p}\right)=\frac{3}{4}\frac{T^{2}}{\epsilon}\left(2\sigma_{\omega}-\frac{3N_{0}^{2}}{\epsilon}\right).
\end{equation}
Then, we can further express $(u\cdot\partial)u^{\nu}$ in Eq.~(\ref{eq:Du_1})
with the results of Eqs.~(\ref{eq:DB_1}) and (\ref{eq:Domega_1}), then
we find, up to $\mathcal{O}(\hbar)$,
\begin{eqnarray}
(u\cdot\partial)u^{\nu} & = & \frac{N_{0}T}{4p}\mathcal{E}^{\nu}+\frac{1}{T}\nabla^{\nu}T\nonumber \\
 &  & +\hbar \frac{1}{96p\pi^{2}}\left(\frac{N_{0}\mu}{2p}(\pi^{2}+\bar{\mu}^{2})-\left(\pi^{2}+3\bar{\mu}^{2}\right)\right)\left[T\epsilon^{\nu\rho\alpha\beta}u_{\rho}(\partial_{\alpha}T)E_{\beta}-\epsilon^{\nu\rho\alpha\beta}u_{\rho}\partial_{\alpha}E_{\beta}\right]\nonumber \\
 &  & -\hbar \frac{N_{0}T^{2}}{96p\pi^{2}}\left[-\frac{1}{2}\frac{N_{0}}{p^{2}}\mu(\pi^{2}+\bar{\mu}^{2})+\frac{3}{4p}(\pi^{2}+3\bar{\mu}^{2})\right]\epsilon^{\nu\rho\alpha\beta}u_{\rho}(\partial_{\alpha}\bar{\mu})E_{\beta}.\label{eq:Du_2}
\end{eqnarray}

Similar, Eq.~(\ref{de0}) and $\partial_{\mu}J^{\mu}=\hbar(\mathbf{E}\cdot\mathbf{B})/(4\pi)$
can be rewritten as, 
\begin{eqnarray}
(\epsilon+p)(3T(u\cdot\partial)\beta+\partial\cdot u)+3N_{0}(u\cdot\partial)\bar{\mu}+\hbar \Pi_{1} & = & 0,\nonumber \\
N_{0}(3T(u\cdot\partial)\beta+\partial\cdot u)+\left.\frac{\partial N_{0}}{\partial(\beta\mu)}\right|_{\beta}(u\cdot\partial)\bar{\mu}+\hbar \Pi_{2} & = & 0,\label{eq:EOM_2}
\end{eqnarray}
where
\begin{eqnarray}
\Pi_{1} & = & -\xi_{B}B^{\mu}(u\cdot\partial)u_{\mu}+\partial_{\mu}(\xi_{B}B^{\mu})-\xi_{\omega}\omega^{\mu}(u\cdot\partial)u_{\mu}+\partial_{\mu}(\xi_{\omega}\omega^{\mu})\nonumber \\
 &  & +\sigma_{B}(E\cdot B)+\sigma_{\omega}(E\cdot\omega),\nonumber \\
\Pi_{2} & = & \partial_{\mu}(\sigma_{B}B^{\mu}+\sigma_{\omega}\omega^{\mu})-CE\cdot B.
\end{eqnarray}
We can solve $(u\cdot\partial)T$ and $(u\cdot\partial)\bar{\mu}$
from Eqs. (\ref{eq:EOM_2}) with $\partial\cdot u=0$, 
\begin{eqnarray}
(u\cdot\partial)T & = & \hbar \frac{2T\sigma_{\omega}\Pi_{1}-3N_{0}T\Pi_{2}}{9N_{0}^{2}-8\epsilon\sigma_{\omega}},\nonumber \\
(u\cdot\partial)\bar{\mu} & = & \hbar \frac{-3N_{0}\Pi_{1}+4\epsilon\Pi_{2}}{T(9N_{0}^{2}-8\epsilon\sigma_{\omega})}.\label{eq:DT_02}
\end{eqnarray}
By using Eqs.~(\ref{eq:DB_1}), (\ref{eq:Domega_1}), and (\ref{eq:Du_2}),
we finally obtain, 
\begin{align}
\Pi_{1} & =(\sigma_{B}-\sigma_{\omega}\frac{N_{0}}{4p})B\cdot\mathcal{E}T+(2\sigma_{\omega}T-\frac{3N_{0}^{2}T}{4p})(\mathcal{E}\cdot\omega),\nonumber \\
\Pi_{2} & =-(CT+\sigma_{B}\frac{N_{0}T}{4p})B\cdot\mathcal{E}+(2\sigma_{B}T-2\sigma_{\omega}\frac{N_{0}T}{4p})(\omega\cdot\mathcal{E}).\label{eq:PI_02}
\end{align}
In the local rest frame, Eqs.~(\ref{eq:Du_2}), (\ref{eq:DT_02}), and (\ref{eq:PI_02})
will reduce to Eqs.~(\ref{UE_UU}), (\ref{TB_TO}), and (\ref{UB_UO}).

\section{Calculations of 2nd-Order Responses in CKT} \label{sec:2ndOrder}
Taking $n^{\mu}=u^{\mu}$ for Eq.~(\ref{global_equil_f}), the local-equilibrium function can be written as 
\begin{eqnarray}
f^\text{eq}_q=f^{(0)}_q+\hbar \delta f^{(\omega)}_q \quad \delta f^{(\omega)}_q=\frac{(\omega\cdot q)}{2(q\cdot u)}\partial_{q\cdot u}f^{(0)}_q,
\end{eqnarray}
which comes from 
\begin{eqnarray}
\frac{\hbar}{2}S^{\mu\nu}_{(u)}\partial_{\mu}(\beta u_{\nu})=
\hbar\beta\frac{(\omega\cdot q)}{2(q\cdot u)}
\end{eqnarray}
based on $S^{\mu\nu}_{(u)}(\sigma_{\mu\nu}+\kappa_{\mu\nu}-T^{-1}u_{\nu}\partial_{\mu}T)=0$. This $\hbar$ correction in $f^\text{eq}_q$ contributes to one of the relevant terms, 
\begin{eqnarray}\nonumber
\hbar q\cdot\Delta\delta f^{(\omega)}_q&=&\hbar q\cdot\Delta\Bigg(\frac{(q\cdot\omega)}{2q\cdot u}\partial_{q\cdot u}f^{(0)}_q\Bigg)
\\\nonumber
&=&-\hbar \frac{(q\cdot\omega)(q\cdot\tilde{E})}{2q\cdot u}\partial^2_{q\cdot u}f^{(0)}_q
+\frac{\hbar}{2q\cdot u}\Bigg[
(q\cdot\partial)(q\cdot\omega)
-\frac{(q\cdot\partial T)(q\cdot\omega)}{T}
-\frac{(q\cdot\omega)}{(q\cdot u)}q^{\mu}\sigma_{\mu\nu}q^{\nu}
\\
&&+\left(\omega^{\alpha}F_{\alpha\beta}q^{\beta}+\frac{(q\cdot\omega)(q\cdot E)}{(q\cdot u)}\right)\Bigg]\partial_{q\cdot u}f^{(0)}_q
.
\end{eqnarray}
On the other hand, the side-jump term yields 
\begin{eqnarray}
\hbar\frac{S^{\mu\nu}_{(u)}E_{\mu}}{(q\cdot u)}\Delta_{\nu}f^{(0)}_q=-\frac{\hbar}{2}\Bigg[(\omega\cdot E)+\frac{(q\cdot E)(q\cdot\omega)}{(q\cdot u)^2}\Bigg]\partial_{q\cdot u}f^{(0)}_q
-\hbar\frac{S^{\mu\nu}_{(u)}E_{\mu}}{(q\cdot u)}\tilde{E}_{\nu}\partial_{q\cdot u}f^{(0)}_q,
\end{eqnarray}
and
\begin{eqnarray}\nonumber
\hbar S^{\mu\nu}_{(u)}(\partial_{\mu}F_{\rho \nu})\partial^{\rho}_{q}f^{(0)}_q&=&\hbar S^{\mu\nu}_{(u)}\Big((\partial_{\mu}E_{\nu})+F^{\rho}_{\mbox{ }\nu}(\partial_{\mu}u_{\rho})\Big)\partial_{q\cdot u}f^{(0)}_q
\\
&=&\hbar\Bigg[ 
S^{\mu\nu}_{(u)}\Big(\partial_{\mu}E_{\nu}+F^{\rho}_{\mbox{ }\nu}\left(\sigma_{\mu\rho}+\kappa_{\mu\rho}\right)\Big)
+\frac{1}{2}\left(E\cdot \omega-\frac{\omega^{\alpha}F_{\alpha\beta}q^{\beta}}{q\cdot u}\right)\Bigg]\partial_{q\cdot u}f^{(0)}_q.
\end{eqnarray}

The derivatives acting on the side-jump term result in
\begin{eqnarray}
\partial_{\rho}S^{\rho\mu}_{(u)}\Delta_{\mu}f^{(0)}_q
&=&
-\frac{\hbar}{2}\Bigg[\omega^{\mu}-\frac{(q\cdot\omega)u^{\mu}}{q\cdot u}
-\frac{(q\cdot\omega)q^{\mu}}{(q\cdot u)^2}
\Bigg]\tilde{E}_{\mu}\partial_{q\cdot u}f^{(0)}_q\notag\\
&&+\frac{\hbar}{2(q\cdot u)}\Bigg[(q\cdot u)\omega^{\mu}(\sigma_{\mu\nu}-\kappa_{\mu\nu})q^{\nu}-(q\cdot\omega)u^{\mu}(\sigma_{\mu\nu}-\kappa_{\mu\nu})q^{\nu}
+\frac{(q\cdot\omega)q^{\mu}\sigma_{\mu\nu}q^{\nu}}{q\cdot u}\Bigg]\partial_{q\cdot u}f^{(0)}_q\notag
\\
&&-\frac{\hbar\epsilon^{\nu\mu\alpha\beta}}{2q\cdot u}\Bigg[q_{\alpha}\kappa_{\mu\nu}
-\frac{u_{\nu}q_{\alpha}q^{\rho}}{q\cdot u}(\sigma_{\mu\rho}+\kappa_{\mu\rho})\Bigg]\tilde{E}_{\beta}\partial_{q\cdot u}f^{(0)}_q.
\end{eqnarray}

Combining all relevant terms above, we derive 
\begin{eqnarray}\nonumber\label{2nd_order_cal}
&&\Bigg[
q\cdot\Delta+\hbar\frac{S_{(u)}^{\mu\nu}E_{\mu}}{(q\cdot u)}\Delta_{\nu}
+\hbar S_{(u)}^{\mu\nu}(\partial_{\mu}F_{\rho \nu})\partial^{\rho}_{q}+\hbar\partial_{\rho}S^{\rho\mu}_{(u)}\Delta_{\mu}\Bigg]f^\text{eq}_q
\\\nonumber
&&=-q\cdot\tilde{E}\partial_{q\cdot u} f^{(0)}_q
+\frac{\hbar}{2}
\Bigg[
\frac{(q\cdot\omega)(q\cdot\tilde{E})}{(q\cdot u)^2}\Big(1-(q\cdot u)\partial_{q\cdot u}\Big)
+\frac{(q\cdot\omega)(u\cdot\partial T)}{T^2}
+\frac{(q\cdot\omega)Tu\cdot\partial\bar{\mu}}{q\cdot u}+\frac{(q\cdot\partial)(q\cdot\omega)}{q\cdot u}
\\\nonumber
&&
\quad
-\omega\cdot\tilde{E}
-\frac{(q\cdot\omega)(u\cdot\partial q\cdot u)}{q\cdot u}
-\frac{(q\cdot\partial T)(q\cdot\omega)}{Tq\cdot u}
+\frac{2S^{\mu\nu}_{(u)}E_{\mu}\tilde{E}_{\nu}}{(q\cdot u)}
+2S^{\mu\nu}_{(u)}\Big(\partial_{\mu}E_{\nu}+F^{\rho}_{\mbox{ }\nu}(\sigma_{\mu\rho}+\kappa_{\mu\rho})\Big)
\\
&&\quad
-\omega^{\mu}(\kappa_{\mu\nu}-\sigma_{\mu\nu})q^{\nu} -\frac{\epsilon^{\nu\mu\alpha\beta}}{q\cdot u}\Bigg(q_{\alpha}\kappa_{\mu\nu}
-\frac{u_{\nu}q_{\alpha}q^{\rho}}{q\cdot u}(\sigma_{\mu\rho}+\kappa_{\mu\rho})\Bigg)\tilde{E}_{\beta}
\Bigg]
\partial_{q\cdot u}f^{(0)}_q
,
\end{eqnarray}
where we use
$u^{\mu}(\kappa_{\mu\nu}-\sigma_{\mu\nu})=0$ and $u^{\mu}(\sigma_{\mu\beta}+\kappa_{\mu\beta})=u\cdot\partial u_{\beta}$.

\section{Anomalous Equation from CKT}
In this appendix, we shall perform the derivation of the anomalous equation in Eq.~(\ref{anomalous_eq}) from CKT by introducing natural boundary conditions at infinity. Based on CKT, the divergence of currents can be written as
\begin{eqnarray}\nonumber\label{divJ_CKT}
\partial_{\mu}J^{\mu}&=&-2\int\frac{d^4q}{(2\pi)^4}\Big(F_{\rho\mu}\partial_q^{\rho}\grave{S}^{<\mu}
-\Sigma^<\cdot \grave{S}^>+\Sigma^>\cdot \grave{S}^<
\Big)
\\\nonumber
&=&-2\int\frac{d^4q}{(2\pi)^3}\bar{\epsilon}(q\cdot n)\Bigg\{F_{\rho\mu}\delta(q^2)\partial^{\rho}_{q}(q^{\mu}f^{(n)}_q)+\hbar F_{\rho\mu}\partial^{\rho}_{q}\Big(\delta(q^2)S^{\mu\nu}_{(n)}\mathcal{D}_{\nu}f^{(n)}_q\Big) 
\\
&&+\hbar\epsilon^{\mu\nu\alpha\beta}F_{\alpha\beta}F_{\rho\mu}\frac{\partial_{q\nu}\delta(q^2)}{4}\partial^{\rho}_qf^{(n)}_q
-\delta(q^2)q\cdot\tilde{\mathcal{C}}-\hbar\epsilon^{\mu\nu\alpha\beta}\mathcal{C}_{\mu}F_{\alpha\beta}\frac{\partial_{q\nu}\delta(q^2)}{4}
\Bigg\},
\end{eqnarray}
where we utilize
\begin{eqnarray}
\frac{\hbar}{4}\epsilon^{\mu\nu\alpha\beta}F_{\alpha\beta}F_{\rho\mu}\partial^{\rho}_q\partial_{q\nu}\delta(q^2)=\frac{\hbar}{2}\epsilon^{\mu\nu\alpha\beta}F_{\alpha\beta}F_{\rho\mu}\partial^{\rho}_q
\left(q_{\nu}\frac{\partial\delta(q^2)}{\partial q^2}\right)=0.
\end{eqnarray}
Here $f^{(n)}_q$ is a general distribution function depending on $q^{\mu}$ and $X^{\mu}$. 
The first classical term corresponding to Lorentz force in Eq.~(\ref{divJ_CKT}) shall vanishes albeit the implicit $\hbar$ corrections encoded in $f^{(n)}_q$, which is due to the fact that this integrand is a total derivative with respect to $q^{\mu}$. Here we introduce the following boundary conditions, 
\begin{eqnarray}
f^{(n)}_q(q_0\rightarrow\infty,{\bf q},X)=0,\quad
f^{(n)}_q(q_0\rightarrow\infty,{\bf q},X)-f^{(n)}_q(q_0\rightarrow-\infty,{\bf q},X)=-1,
\end{eqnarray}
where $q_0=q\cdot n$. 
The conditions assume the vanishing distribution functions at infinity, while the $-1$ in the second equality above comes from the anti-commutation relation for fermions. By using such boundary conditions, one can explicitly show
\begin{eqnarray}
-2\int\frac{d^4q}{(2\pi)^3}\bar{\epsilon}(q\cdot n)F_{\rho\mu}\delta(q^2)\partial^{\rho}_{q}(q^{\mu}f^{(n)}_q)=0.
\end{eqnarray}  

Next, we consider the side-jump term, which yields 
\begin{eqnarray}\nonumber
&&-2\int\frac{d^4q}{(2\pi)^3}\bar{\epsilon}(q\cdot n)\hbar F_{\rho\mu}\partial^{\rho}_{q}\Big(\delta(q^2)S^{\mu\nu}_{(n)}\mathcal{D}_{\nu}f^{(n)}_q\Big)
\\\nonumber
&&=\hbar\int\frac{d^3{\bf q}}{(2\pi)^3} \frac{({\bf E\times q})}{2}\cdot\Bigg[\frac{\bar{\epsilon}(q_0)}{|q_0|}\bm{\mathcal{D}}f^{(n)}_q(q_0,q_i)\Bigg]^{q_0=\infty}_{q_0=-\infty}
\\\nonumber
&&\quad
-\hbar\int\frac{dq_0d^2{\bf q}_{\perp}}{(2\pi)^3}\bar{\epsilon}(q_0)
\Bigg[\frac{\delta(q^2)}{q_0}\Big(q_{\parallel}({\bf B}\cdot\bm{\mathcal{D}})-({\bf q\cdot B})\mathcal{D}_{\parallel}\Big)f^{(n)}_q\Bigg]^{q_{\parallel}=\infty}_{q_{\parallel}=-\infty} 
\\
&&=0,
\end{eqnarray}
where ${\bf q}_{\perp}$ denotes the spatial momentum perpendicular to ${\bf q}_{\parallel}$ in the second integral above and we introduce the boundary condition
\begin{eqnarray}
(\mathcal{D}_{\mu}f^{(n)}_q)_{q_0=\infty}=(\mathcal{D}_{\mu}f^{(n)}_q)_{q_0=-\infty}=0
\end{eqnarray}
such that the derivatives of distribution functions for (anti-)particles in phase space and collisions vanish at infinity. 

Finally, using the integration by part and subsequently performing the integration with $q_0$, the $\partial_{q\nu}\delta(q^2)$ term gives
\begin{eqnarray}\nonumber\label{chiral_anomaly}
&&-2\int\frac{d^4q}{(2\pi)^3}\bar{\epsilon}(q\cdot n)\hbar\epsilon^{\mu\nu\alpha\beta}F_{\alpha\beta}F_{\rho\mu}\frac{\partial_{q\nu}\delta(q^2)}{4}\partial^{\rho}_qf^{(n)}_q
\\
&&=\hbar({\bf E\cdot B})\int\frac{d^3{\bf q}}{2(2\pi)^3|{\bf q}|}
(\partial^2_{|{\bf q}|}-\nabla^2_{\bf q})\Big(f^{(n)}_{\bf q}(|{\bf q}|,{\bf q})-f^{(n)}_{\bf q}(-|{\bf q}|,{\bf q})\Big),
\end{eqnarray}
where $f_{\bf q}$ as a function of ${\bf q}$ denotes the distribution function with the on-shell condition and ${\bf E}$ and ${\bf B}$ are defined in the $n^{\mu}$ frame. Here $\pm|{\bf q}|$ in the parentheses of $f^{(n)}_{\bf q}$ distinguish the particles and anti-particles. 
Note that here $\partial_{|{\bf q}|}$ and $\nabla_{\bf q}$ are partial derivatives. We should apply the chain rule to rewrite them in terms of total derivatives. By utilizing
\begin{eqnarray}
\frac{d}{d|{\bf q}|}f(|{\bf q}|, {\bf q})=\big(\partial_{|{\bf q}|}+\hat{\bf q}\cdot\nabla_{\bf q}\big)f(|{\bf q}|, {\bf q}),\quad
\frac{d}{d{\bf q}}f(|{\bf q}|, {\bf q})=\big(\nabla_{\bf q}+\hat{\bf q}\partial_{|{\bf q}|}\big)f(|{\bf q}|, {\bf q}),
\end{eqnarray}
where $\hat{\bf q}={\bf q}/|{\bf q}|$,
we obtain
\begin{eqnarray}\nonumber
&&\int\frac{d^3{\bf q}}{2(2\pi)^3|{\bf q}|}
(\partial^2_{|{\bf q}|}-\nabla^2_{\bf q})\Big(f^{(n)}_{\bf q}(|{\bf q}|,{\bf q})-f^{(n)}_{\bf q}(-|{\bf q}|,{\bf q})\Big)
\\\nonumber
&&=\int\frac{d^3{\bf q}}{2(2\pi)^3|{\bf q}|}
\Bigg(\left(\frac{d}{d|{\bf q}|}-\hat{\bf q}\cdot\nabla_{\bf q}\right)\partial_{|{\bf q}|}-\left(\frac{d}{d{\bf q}}-\hat{\bf q}\partial_{|{\bf q}|}\right)\cdot\nabla_{\bf q}\Bigg)\Big(f^{(n)}_{\bf q}(|{\bf q}|,{\bf q})-f^{(n)}_{\bf q}(-|{\bf q}|,{\bf q})\Big)
\\
&&=\int\frac{d^3{\bf q}}{2(2\pi)^3|{\bf q}|}
\Bigg(\frac{d}{d|{\bf q}|}\partial_{|{\bf q}|}-\frac{d}{d{\bf q}}\cdot\nabla_{\bf q}\Bigg)\Big(f^{(n)}_{\bf q}(|{\bf q}|,{\bf q})-f^{(n)}_{\bf q}(-|{\bf q}|,{\bf q})\Big).
\end{eqnarray}
Next, we implement the integration by part and find that the integral becomes
\begin{eqnarray}\nonumber
&&\int\frac{d^3{\bf q}}{2(2\pi)^3|{\bf q}|}
\Bigg(\frac{d}{d|{\bf q}|}\partial_{|{\bf q}|}-\frac{d}{d{\bf q}}\cdot\nabla_{\bf q}\Bigg)\Big(f^{(n)}_{\bf q}(|{\bf q}|,{\bf q})-f^{(n)}_{\bf q}(-|{\bf q}|,{\bf q})\Big)
\\\nonumber
&&=-\int\frac{d^3{\bf q}}{2(2\pi)^3|{\bf q}|^2}
\Bigg(\partial_{|{\bf q}|}+\hat{\bf q}\cdot\nabla_{\bf q}\Bigg)\Big(f^{(n)}_{\bf q}(|{\bf q}|,{\bf q})-f^{(n)}_{\bf q}(-|{\bf q}|,{\bf q})\Big)
\\
&&=-\int\frac{d^3{\bf q}}{2(2\pi)^3|{\bf q}|^2}
\frac{d}{d|{\bf q}|}\Big(f^{(n)}_{\bf q}(|{\bf q}|,{\bf q})-f^{(n)}_{\bf q}(-|{\bf q}|,{\bf q})\Big),
\end{eqnarray}
which turns out to be a surface term. We thus derive
\begin{eqnarray}\nonumber
&&\int\frac{d^3{\bf q}}{2(2\pi)^3|{\bf q}|}
(\partial^2_{|{\bf q}|}-\nabla^2_{\bf q})\Big(f^{(n)}_{\bf q}(|{\bf q}|,{\bf q})-f^{(n)}_{\bf q}(-|{\bf q}|,{\bf q})\Big)
\\\nonumber
&&=-\int\frac{d\Omega}{2(2\pi)^3}\Bigg[f^{(n)}_{\bf q}(|{\bf q}|,{\bf q})-f^{(n)}_{\bf q}(-|{\bf q}|,{\bf q})\Bigg]^{|{\bf q}|\rightarrow\infty}_{|{\bf q}|\rightarrow 0}
\\
&&=\frac{1}{4\pi^2},
\end{eqnarray}
which results in the chiral anomaly
\begin{eqnarray}
-2\int\frac{d^4q}{(2\pi)^3}\bar{\epsilon}(q\cdot n)\hbar\epsilon^{\mu\nu\alpha\beta}F_{\alpha\beta}F_{\rho\mu}\frac{\partial_{q\nu}\delta(q^2)}{4}\partial^{\rho}_qf_q=\frac{\hbar}{4\pi^2}({\bf E\cdot B}).
\end{eqnarray}
The divergence of currents then reads
\begin{eqnarray}
\partial_{\mu}J^{\mu}=\frac{\hbar}{4\pi^2}({\bf E\cdot B})+2\int\frac{d^4q}{(2\pi)^3}\bar{\epsilon}(q\cdot n)\left[\delta(q^2)q\cdot\tilde{\mathcal{C}}+\hbar\epsilon^{\mu\nu\alpha\beta}\mathcal{C}_{\mu}F_{\alpha\beta}\frac{\partial_{q\nu}\delta(q^2)}{4}\right].
\end{eqnarray}
In realistic cases, the integral with collisional terms should vanish and the divergence of currents gives rise to the well-known anomalous equation.

\acknowledgments
Y. H. was partially supported by Japan Society of Promotion of Science (JSPS), Grants-in-Aid for Scientific Research
(KAKENHI) Grants No. 15H03652, 16K17716, and 17H06462.  Y. H. was also partially supported by RIKEN iTHES Project and iTHEMS Program.
S. P. is supported by a JSPS postdoctoral fellowship for foreign researchers Grant No. JP16F16320. D. Y. was supported by the RIKEN Foreign Postdoctoral Researcher program.

\bibliography{nonlinear_effects_CKT.bbl}

\end{document}